\documentclass[12pt, preprint]{aastex}

\begin{document}

 \title{The Stability of Double White Dwarf Binaries Undergoing Direct Impact  Accretion}

 \author{Patrick M.\ Motl, Juhan Frank, Joel E.\ Tohline \& Mario C.\ R.\ D'Souza}
 \affil{Department of Physics \& Astronomy,
          Louisiana State University,
          Baton Rouge, LA 70803}

 \begin{abstract}
 We present numerical simulations of dynamically unstable mass transfer in a double white
 dwarf binary with initial mass ratio, $q = 0.4$.  The binary components are approximated
 as polytropes of index $n = 3/2$ and the initially synchronously rotating,  semi-detached equilibrium
 binary is evolved hydrodynamically with the gravitational potential being computed through
 the solution of Poisson's equation.  Upon initiating deep contact
 in our baseline simulation,
 the mass transfer rate
 grows by more than an order of magnitude over approximately ten orbits, as would be expected
 for dynamically unstable mass transfer.  However, the mass transfer rate then  reaches a
 peak value, the binary expands and the mass transfer event subsides.  The binary must
 therefore have crossed the critical mass ratio for stability against dynamical mass transfer.
 Despite the initial loss of orbital  angular momentum into the spin of the accreting star,
 we find that the accretor's spin saturates and angular momentum is returned to the orbit more
 efficiently than has been previously suspected for binaries in the direct impact accretion mode.
 To explore this surprising result, we directly measure the critical mass ratio for stability by
 imposing artificial angular momentum loss at various rates to drive the binary to an equilibrium
 mass transfer rate.  For one of these driven evolutions, we attain equilibrium mass transfer
 and deduce that effectively $q_{\rm crit}$ has evolved to approximately $2/3$. Despite the absence of
 a fully developed disk, tidal interactions appear effective in returning excess spin
 angular momentum to the orbit. 
  \end{abstract}


 \section{Introduction}

Binaries containing compact components are of great current interest in astrophysics. For an
up-to-date online review of the origin and evolution of binaries with all types of compact components
see \citet{PY06}. More specifically, for recent discussions of the orbital evolution and stability
of mass transfer in the case of double white dwarf binaries, see \citet{Maet04} and
\citet{GPF} (GPF).
 White dwarf stellar remnants represent the most frequent endpoint for stellar evolution
and are therefore the most common constituents of binaries with compact components.
When  common envelope evolution occurs, a white dwarf pair can emerge as a sufficiently
compact binary for gravitational radiation to drive the binary to a semi-detached state over
an astrophysically interesting timescale.  It has been estimated that
$\sim 1.5\times 10^8$
double white dwarf (DWD) binaries exist within the Galaxy, of which approximately 1/3 are semi-detached
\citep{NYPZ01}.  The richness of these
binaries  will in fact impose a confusion-limited
background for low-frequency gravitational wave detectors such as LISA
\citep{HBW90, HiBe00, NYPZ01}.
In this paper, we focus on the behavior of such
double white dwarf binaries from the point where mass transfer ensues by evolving
an equilibrium, semi-detached polytropic binary hydrodynamically.
We leave the complex and interesting phenomena that bring such binaries into contact
for other studies.

For a cool white dwarf star with mass well below the  Chandrasekhar mass,  gravity is balanced by
the degeneracy pressure of non-relativistic electrons.
For this study, we use a polytropic equation of state as an approximation to an
idealized, cool white dwarf.
For a polytrope with index $n = \frac{3}{2}$ and corresponding adiabatic exponent
$\gamma = \frac{5}{3}$, the radius--mass relation is
\begin{equation}
  R \propto M^{- 1/3}.
  \label{mr}
\end{equation}
If both components of a DWD binary have
equal entropy  (equal polytropic constants in  the view of this approximate
equation of state), it must be the less massive star that will reach contact with its
Roche lobe first.  Furthermore, Equation  (\ref{mr})
 indicates that the donor will expand upon mass loss
to the accretor.
This fact introduces the possibility that the resulting mass transfer may be dynamically unstable
as the donor can expand well beyond its Roche lobe leading to catastrophic mass transfer.

If we neglect the angular momentum in the spin of the binary components and allow
all the angular momentum contained in the mass transfer stream to be returned to
the orbit by tides, we would expect that
a binary composed of stars obeying the mass-radius relation in Equation (\ref{mr})
 is stable provided that \citep{P67}
\begin{equation}
   q = \frac{M_{D}}{M_{A}} \leq \frac{2}{3} = q_{\rm crit},
   \label{qdisk}
\end{equation}
 where $M_{D}$ and $M_{A}$ are, respectively, the masses of the donor and accretor.  However, it is now
 well known that the exchange of angular momentum in DWD binaries dramatically alters
 their stability properties (Marsh \textit{et al.}\ 2004; GPF).
 In many instances, the accretion stream will directly strike the
 surface of the accretor instead of orbiting to eventually form an accretion disk.  As is well-known, a
 tidally truncated accretion disk can efficiently return the advected angular momentum to
 the orbit.  However, in the direct impact accretion mode the stream spins up the
 accretor and it is not clear how much angular momentum can be returned to the orbit. 
 We note that direct impact accretion is of relevance in cases where the binary components
 are of comparable size.  In addition to DWD binaries, similar circumstances arise in the
 evolution of Algol binaries as well, for example.

 The importance of the fundamental uncertainty in the efficiency of tidal coupling
 was highlighted in the population synthesis calculations of \citet{NZVY01}.
 In the idealized case where no angular momentum is lost in the
 accretor's spin (as would hold for the prediction  in Equation \ref{qdisk}),
 approximately $20\%$ of DWD binaries can survive
 mass transfer to become long-lived, AM CVn-type binaries.  On the other hand, if none of the
 accretor's spin angular momentum is returned to the orbit, the effective mass ratio for stable
 mass transfer is reduced significantly from the expectation reflected in Equation (\ref{qdisk})
 %-
  and very few AM CVn can arise from a DWD progenitor.
 %-
 Rather, most DWD binaries would be disrupted
 shortly after becoming semi-detached, followed by a disk or common envelope phase and
 ending in merger.  In the event that the total system mass exceeds the
 Chandrasekhar limit, the catastrophic merger would result in a, perhaps anomalous, type
 Ia supernova.

 We present here a series of evolutions of an initially semi-detached DWD binary model
 with an initial mass ratio of $q = 0.4$.   The accretor is sufficiently large  for
 the resulting mass transfer  stream to strike the accretor directly.
 This binary thus begins its evolution as a stable system
 with respect to mass transfer in the limiting case of efficient tidal coupling but will be
 unstable if
 direct impact accretion removes angular momentum from the binary orbit.
In \S 2, we present our baseline evolution of this model through 43 orbits and demonstrate
that while mass transfer does grow rapidly in the initial phase, the binary separates and the
mass transfer rate then declines allowing the binary to survive.  To further investigate this result,
we measure the mass ratio for stability to dynamical mass transfer in \S 3 by repeating  our
initial simulation with various artificial driving rates of angular momentum loss.
We find that, effectively, this model binary transitions from unstable mass-transfer to
stable mass transfer; at late times it appears to behave
like a system that returns angular momentum rapidly to the binary orbit despite the absence of
a tidally truncated accretion disk.
In \S 4, we present a simple analysis
for understanding these evolutions in terms of orbit-averaged equations.  This analysis demonstrates
 that, in our simulations, the rate of change in the spin of the accretor saturates and subsequently
 the mass transfer rate begins to subside.  In the concluding \S 5, we summarize our results
 and highlight some important points that may limit the application of our results to real DWD
 binaries.

 \section{Baseline Evolution of the Q0.4 Binary}

We begin by considering the numerical evolution of a polytropic binary with an initial mass
ratio of $q = 0.4$.
The computational tools for performing this simulation and for generating the initial data
have been described previously in \citet{MTF} (MTF) and include the important
modifications described in \citet{DMTF} (DMTF) that enable long-term evolutions of
binary models.   In the evolution, we solve the equations for ideal, Newtonian hydrodynamics
while simultaneously solving Poisson's equation to give a consistent gravitational potential
at every timestep in the evolution.

The initial data describe an $n = \frac{3}{2}$ polytropic binary that is synchronously rotating
on a circular orbit.  The relevant parameters for the $q = 0.4$ (hereafter Q0.4) model are
listed in Table 1.   As in DMTF, we use polytropic units where the radial extent of the
computational grid for the SCF model, the maximum density of one binary component
and the gravitational constant are all unity.
In Table 1, we list initial values for the mass ratio $q_{0}$, separation $a_{0}$, angular
frequency $\Omega_{0}$, total angular momentum $J^{\rm tot}_{0}$ and the offset $R_{\rm com}$ between
the system center of mass and the origin of the evolution grid (which is much less than
the grid spacing used $\Delta R = \frac{1}{127} \approx 7.87 \times 10^{-3}$).  We also
have tabulated  values for the component's mass $M_{i}$, maximum density
$\rho^{\rm max}_{i}$, polytropic constant $\kappa_{i}$, stellar volume $V_{i}$ and Roche
lobe volume $V^{RL}_{i}$.

Note in particular
that the polytropic constants, $\kappa_{i}$, are not equal for the two components and that
the volume of the more massive star, ultimately the accretor, exceeds the volume of the
donor which is nearly in contact with its Roche lobe.
In the formulation of the SCF code
we have used to create this initial data (see MTF for details),
we must specify parameters that do not
directly correspond to a desired binary  with a prescribed mass or entropy
ratio.  The Q0.4 model was generated by trial and error and while it posses an interesting
mass ratio, its components do not conform to the desired mass-radius relation of
Equation (\ref{mr}) relative to
one another due to the difference in polytropic constants.  We have begun to develop an improved
SCF technique that is capable of enforcing a desired mass ratio and mass-radius relation
\textit{a priori} but for now we proceed with the Q0.4 model as it stands.  Though the accretor is
well inside its Roche lobe initially, and becomes even more confined in its potential well
as the evolution proceeds, the reader is encouraged to be mindful of the caveat that the accretor's
size may accentuate gravitational torques on the binary leading the system to be more
stable to dynamical mass transfer than real DWD binaries of the same initial mass ratio.

To make this cautionary note concrete, from Table 1 one can conclude that the effective 
radius of the accretor is approximately 1.6 times larger than it should be based upon 
the mass radius relation in Eq. (1) due to the differing polytropic constants of the 
two components.  The magnitude of the tidal field of the accretor should scale 
as $\left(R_{A} / a\right)^{6}$ (see for example \citet{C84} or \citet{Z05}).  
The accretor's tidal field in this specific binary model is thus overestimated by 
a factor of approximately 17.

\clearpage
\begin{table}\caption{Initial Parameters for Model Q0.4\tablenotemark{\dag}}
   \begin{center}
      \begin{tabular}{lc|lcc}
         \hline
         \hline
         System & Initial SCF & Component & & \\
          Parameter &  Value &  Parameter & Donor & Accretor \\
         \hline
         $q_{0}$ & 0.4085 & $M_{i}$ & $6.957 \times 10^{-3}$ & $1.703 \times 10^{-2}$ \\
         $a_{0}$ & 0.8169 & $\rho^{\rm max}_{i}$ & 0.71 & 1.0 \\
         $\Omega_{0}$ & 0.2112 & $\kappa_{i}$ & $1.904 \times 10^{-2}$ & $3.119 \times 10^{-2}$ \\
         $J^{\rm tot}_{0}$ & $7.794 \times 10^{-4}$ &  $V_{i}$ & $6.180 \times 10^{-2}$ & 0.1041 \\
         $R_{\rm com}$ & $2.669 \times 10^{-6}$ & $V^{RL}_{i}$ & $6.204 \times 10^{-2}$ & 0.2154 \\
         \hline
      \end{tabular}
   \end{center}
   \tablenotetext{\dag}{See MTF and DMTF for further details about these parameters.}
\end{table}
\clearpage

The initial binary model described above is introduced into the hydrodynamics code
and evolved on a cylindrical mesh with 162 radial by 98 vertical by 256 azimuthal zones.
The evolution is performed in a rotating frame of reference where the frame frequency is
taken to be the angular frequency consistent with the initial data, $\Omega_{0}$ and this frame
frequency is kept fixed during the evolution.  The center of mass correction described
in DMTF is used and we drain angular momentum from the binary to drive the donor into
deep contact with its Roche lobe.  Specifically, we remove angular momentum
as described in DMTF at a rate of
$1\%$ per orbit for the first 1.6 orbits where the orbital time, $P_{0}=2\pi/\Omega_0$.
The initial driving tightens the binary, causing a
well-resolved accretion stream to form.  This speeds up the numerical evolution and in the
event of dynamically unstable mass transfer, the mass transfer rate will rapidly grow to a large
amplitude in any event.

\clearpage
\begin{figure}
   \begin{center}
      \begin{tabular}{cc}
         \includegraphics[scale=0.75]{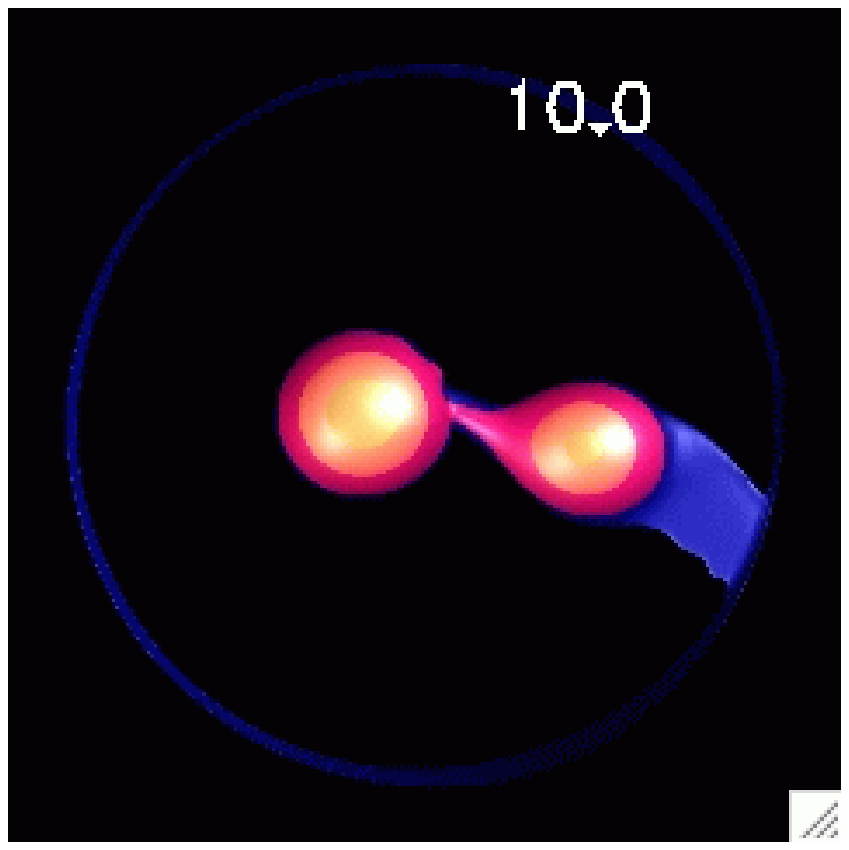} &
         \includegraphics[scale=0.75]{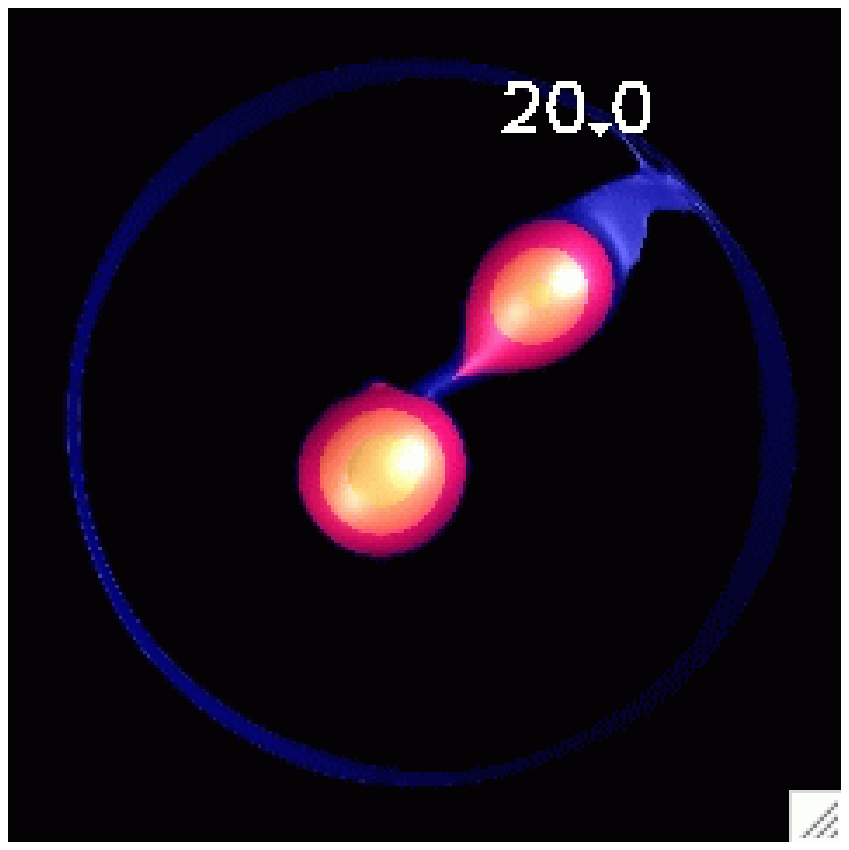} \\
         \includegraphics[scale=0.75]{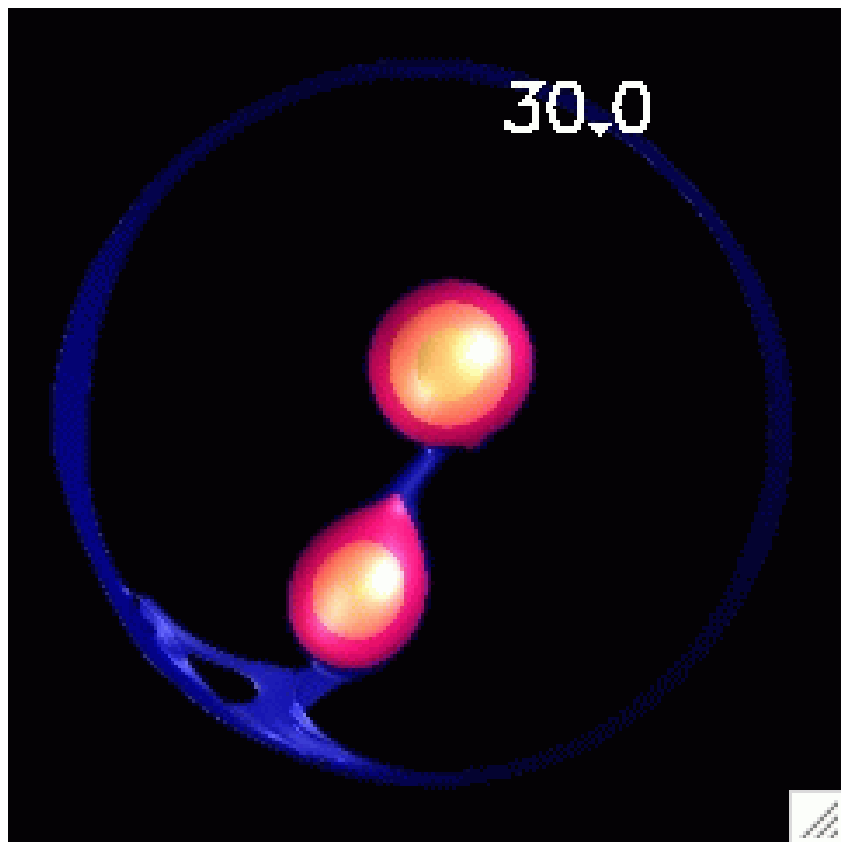} &
         \includegraphics[scale=0.75]{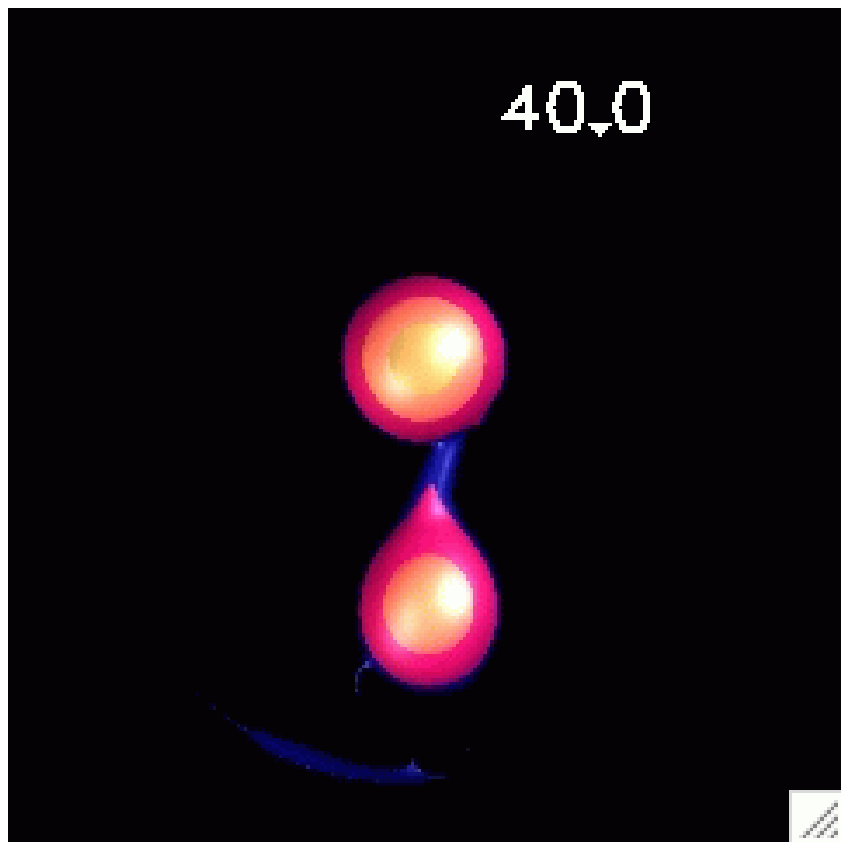}  \\
      \end{tabular}
   \end{center}
   \caption{Four still frames from the baseline evolution of the Q0.4 binary, viewed from above
   along the rotational axis.  The images show, starting in the upper left hand corner, the binary
   at 10, 20, 30 and 40 times the initial binary period.  The innermost shell depicts a surface of
   constant density at $0.5 \; \rho_{\rm max}$ while the yellow, red and blue surfaces correspond to
   density levels of $0.1, 10^{-3}$ and $10^{-5} \;  \rho_{\rm max}$.  In the online version, we present
   an MPEG movie of the entire evolution through 43 orbits.}
   \label{q04a_movie}
\end{figure}
\clearpage

 Images from the baseline Q0.4 evolution are shown in Figure \ref{q04a_movie}
 at 10, 20, 30 and 40 $P_{0}$.
The binary is viewed from above with counter-clockwise motion corresponding to the
binary advancing in the, initially, co-rotating frame.  The different colored shells are isopycnic
surfaces corresponding to density levels of $0.5, 0.1, 10^{-3}$ and $10^{-5}$ of the maximum
density from the innermost to outermost, blue, surface.  At 10 $P_{0}$,  the mass transfer rate
is near its peak value, corresponding to approximately $0.4\%$ of the donor's initial mass
per orbit.  In the subsequent frames, the binary separates (moves clockwise about the rotational
axis), the mass transfer stream diminishes and the binary attains stable, long-lived mass
transfer with a rather gentle evolution.

Two numerical artifacts are apparent in these images and warrant further discussion.  In addition
to the mass transfer stream at L1, there is an outflow of low density material from the backside
of the donor.  This outflow is a numerical error and its presence establishes a floor level for the
mass transfer rate that is resolved and corresponds to physical mass loss from the donor
(see DMTF for further discussion).  In our baseline simulation, the mass flux at L1 dominates this
numerical effect by over an order of magnitude.  Integrated over the entire evolution, less that
0.1\% of the initial system mass exits the computational domain through this outflow.  In addition,
we enforce a vacuum level on the mass density of $10^{-10}$ for numerical stability.  The
vacuum level should be compared to the maximum density of order 1 and the characteristic
density at the edge of the stars which is approximately $10^{-5}$.  Enforcing this vacuum level
adds mass to the grid at a rate nearly equal to the outflow with the net effect that the total
mass increases by $4 \times 10^{-6}$ relative to its initial value over the baseline
evolution.
We also wish to discuss the ring of material that forms when the donor's outflow interacts
with the grid boundary.  We implement simple outflow boundary conditions (MTF) that
are mathematically consistent only in the case where the fluid velocity normal to the
boundary is supersonic (please see Stone and Norman 1992 for further discussion
of boundary conditions for Eulerian codes).  Using our simple formulation of the
boundary conditions as opposed to a characteristic treatment allows material to
accumulate against the grid edge but its presence does not appear to influence
our results.

\begin{figure}
   \begin{center}
       \begin{tabular}{cc}
          \includegraphics[scale=0.4]{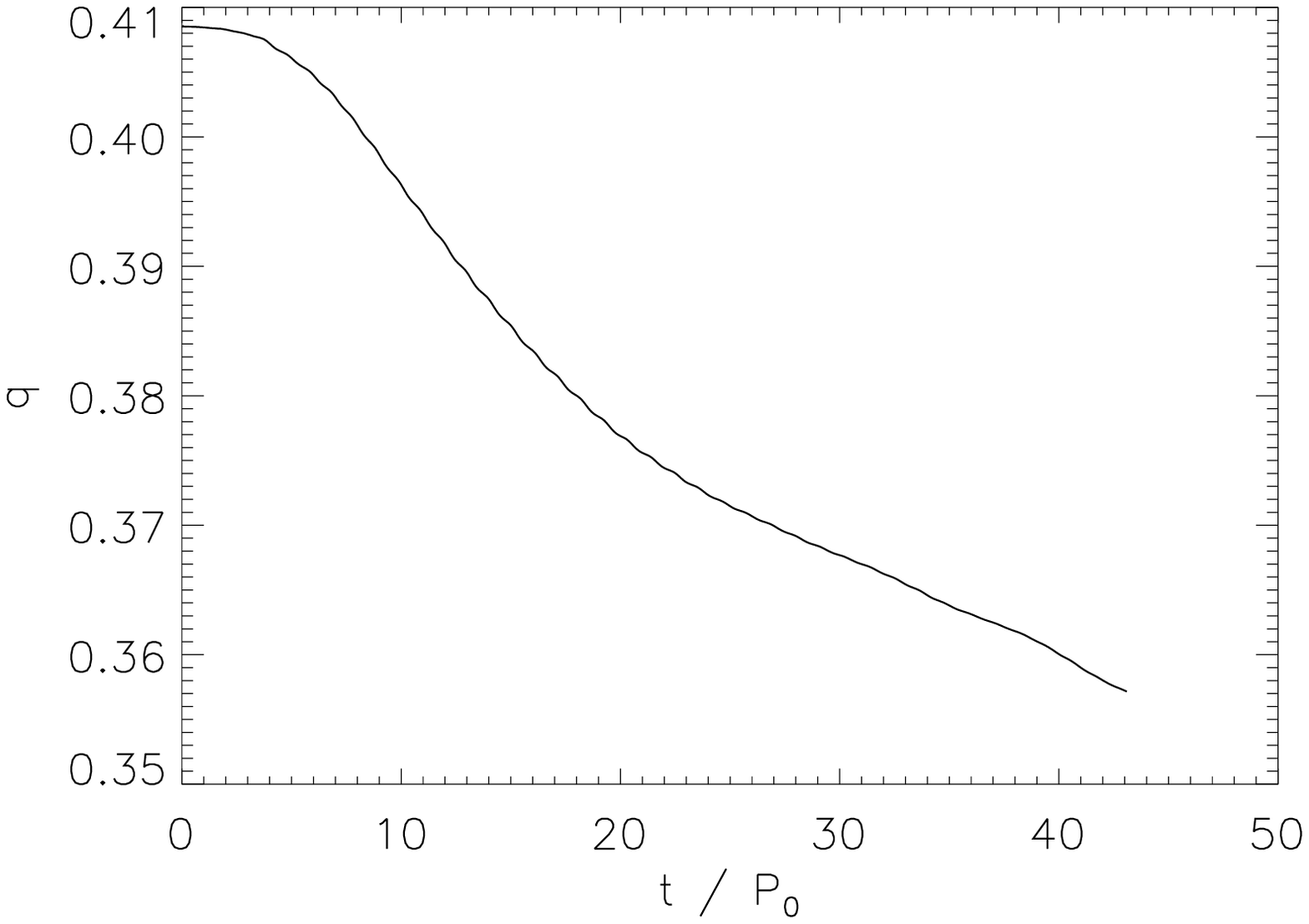} &
          \includegraphics[scale=0.4]{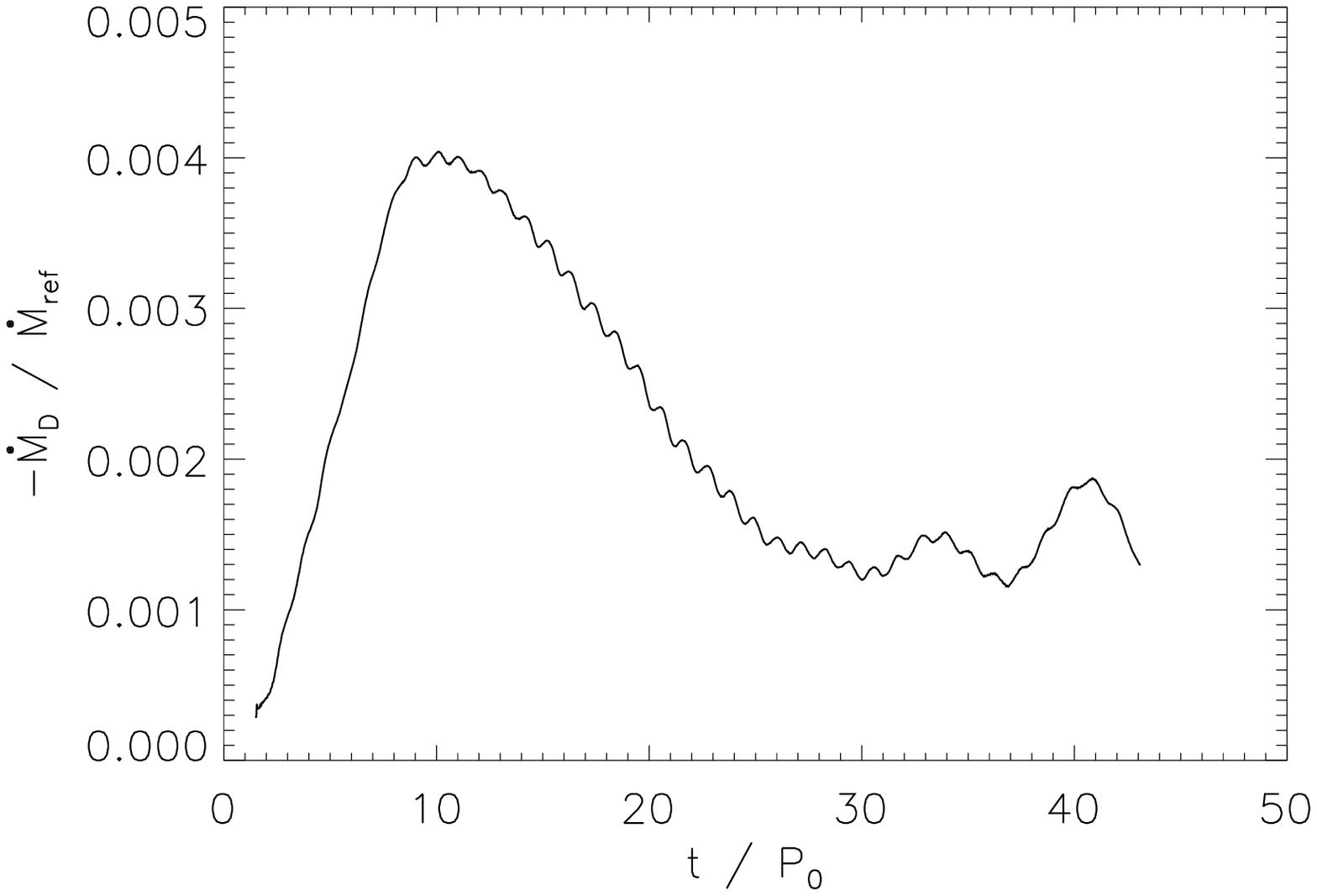} \\
          \includegraphics[scale=0.4]{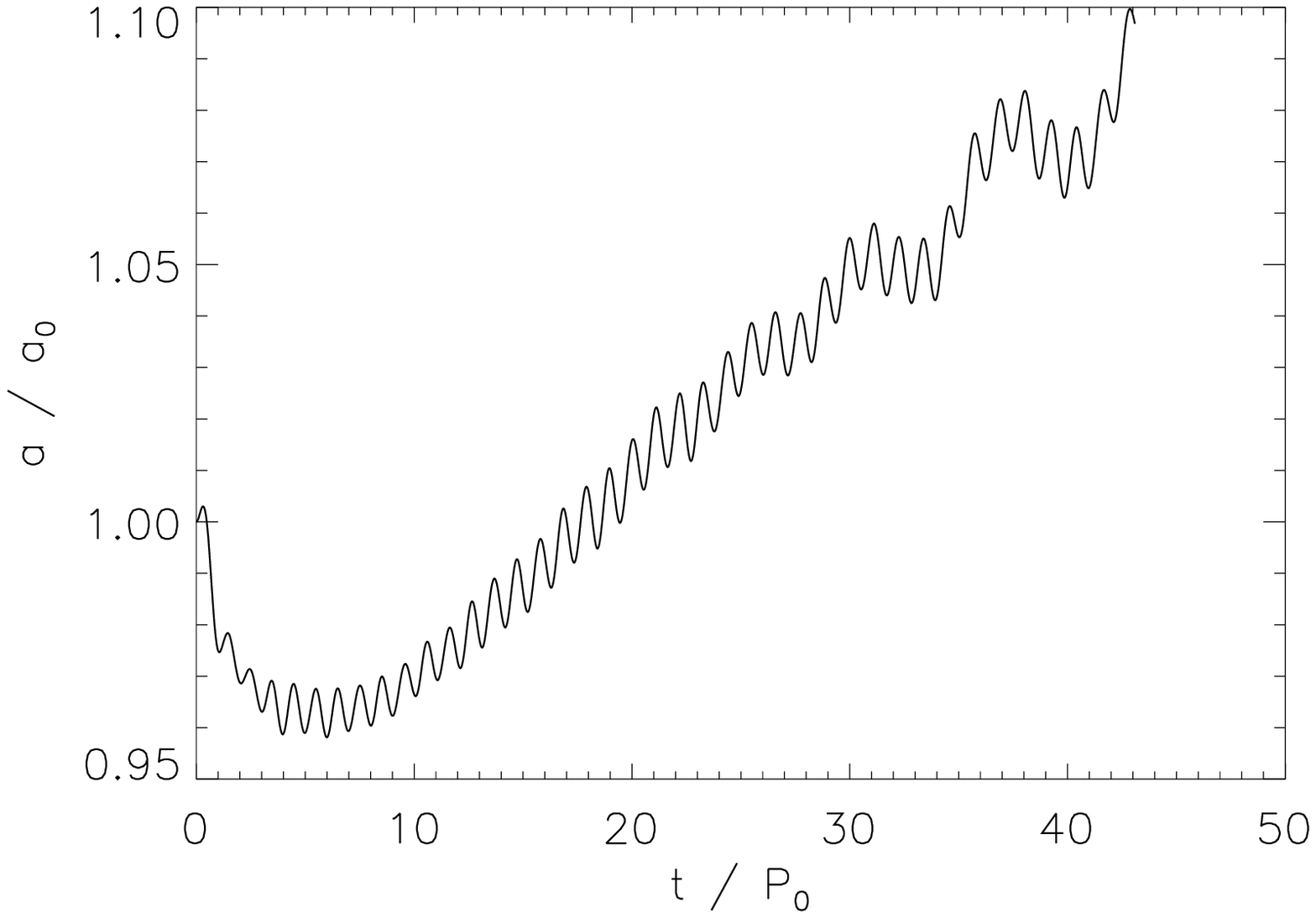} &
          \includegraphics[scale=0.4]{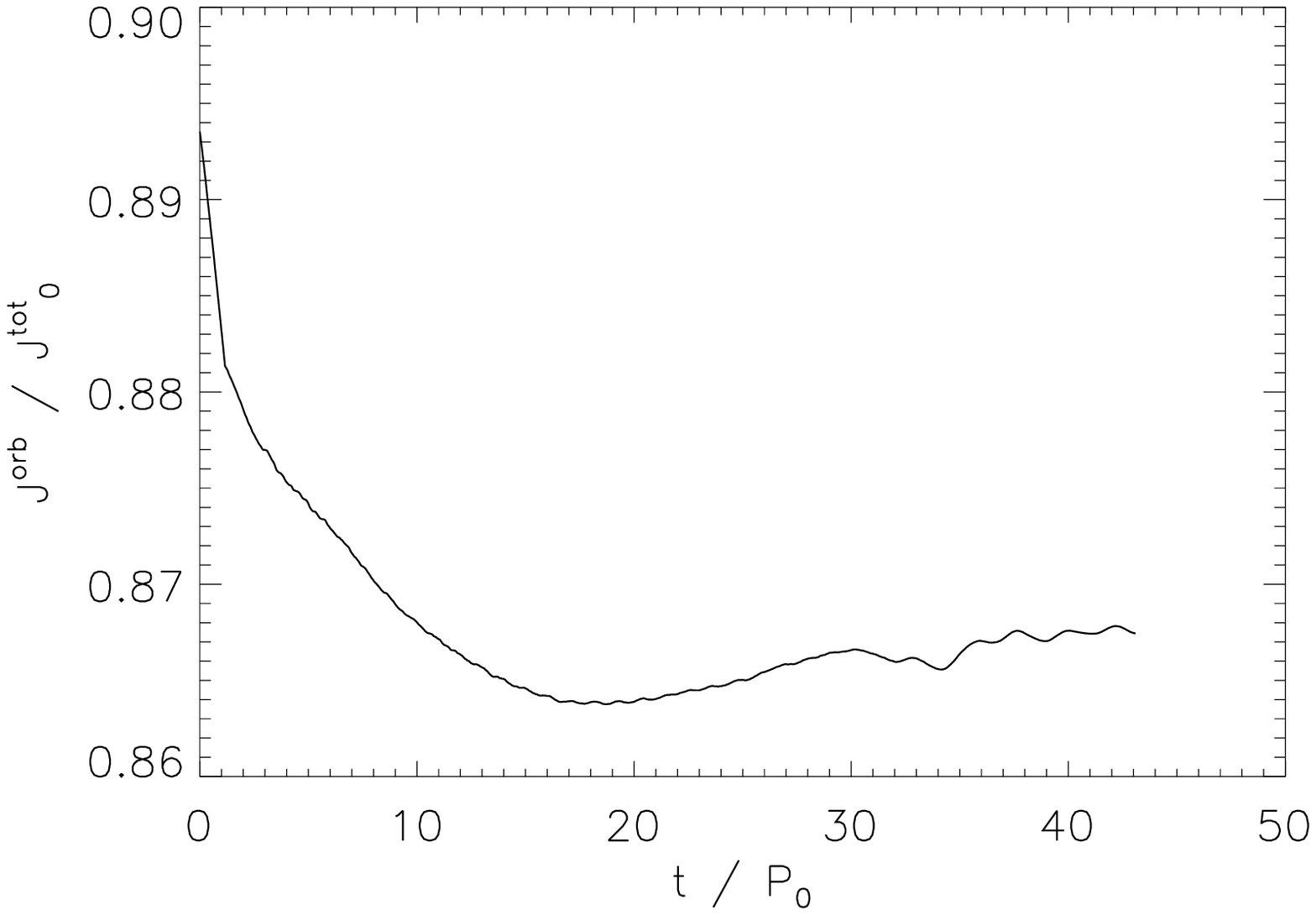} \\
          \includegraphics[scale=0.4]{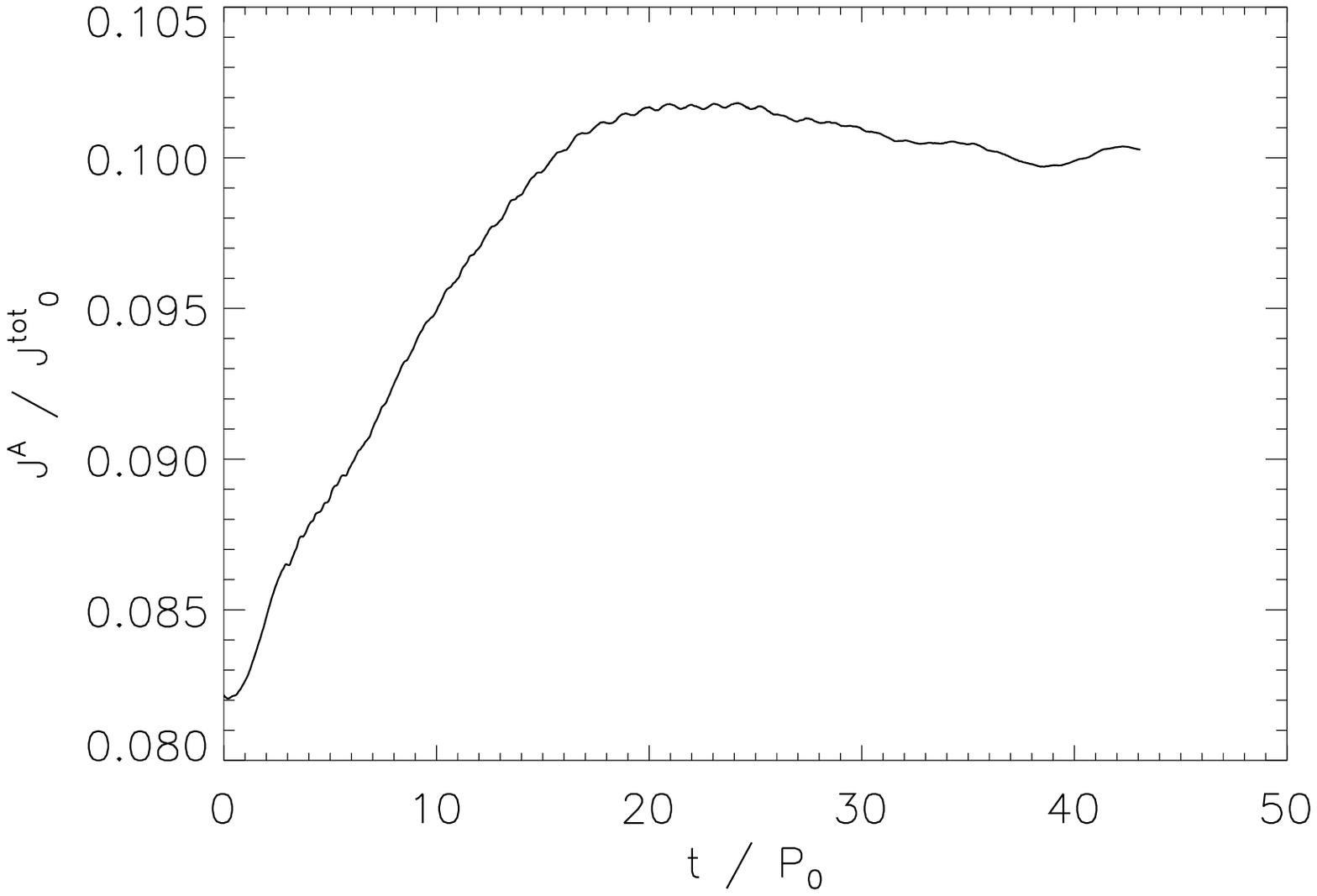} &
          \includegraphics[scale=0.4]{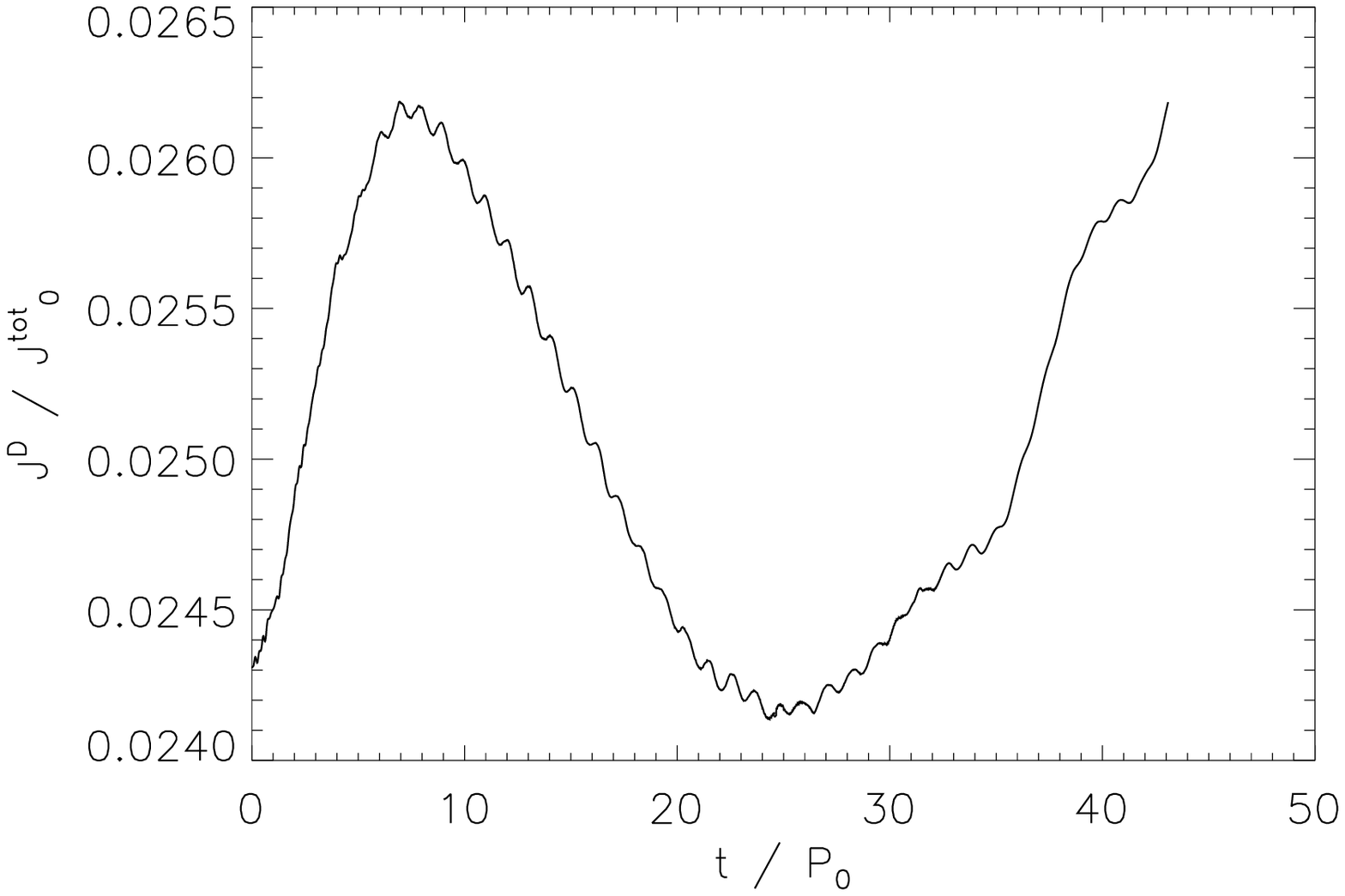} \\
       \end{tabular}
    \end{center}
    \caption{Relevant binary parameters through the baseline evolution
    of the Q0.4 system; time is shown in units of $P_0$.
    Beginning in the upper left hand panel, we show the binary
    mass ratio $q = M_{D} / M_{A}$.  The mass transfer rate, normalized to
    $\dot{M}_{\rm ref} = M_{D} / P_{0}$ is shown in the upper right hand panel.
    The mass transfer rate has been averaged with box-car smoothing over 3 orbital
    period intervals.  In the middle row we show the orbital separation,
    normalized to its initial value, and the orbital angular momentum, normalized to the
    system's initial, total angular momentum $J_0^{\rm tot}$.
    In the bottom row we show the spin angular momentum of the accretor (left panel)
    and donor (right panel) where each has been normalized by $J_0^{\rm tot}$.}
    \label{q04a_plots}
 \end{figure}

We examine the baseline evolution in more detail with the plots shown in Figure \ref{q04a_plots}.
The mass ratio decreases throughout the simulation, falling by $\sim 12\%$ through the 43 orbits of
the evolution.
The smoothed mass transfer rate is shown in the top right panel of Figure  \ref{q04a_plots}.
The donor loses mass at
a rapidly accelerating rate over the first 10 orbits of the evolution, the mass-transfer rate then
peaks and begins to diminish to a nearly constant level from about 25 orbits through the
end of the evolution.  Modulations in the orbit, reflected in a rather complicated pattern
in the ellipticity of the nearly circular orbit, result in a varying depth of contact and hence
a varying mass transfer rate in the last quarter of the evolution.  The orbital separation is shown
in the middle left panel of Figure \ref{q04a_plots}.
The initial driving causes the binary to contract but  soon the
binary begins to expand and the expansion continues throughout the run.
The baseline simulation is stopped after 43 orbits
because the binary has separated to the point that the
components approach the boundary of the computational domain.

\clearpage
\clearpage

 To complete the description of the baseline evolution, we also plot several angular momentum
 components of the binary system in Figure \ref{q04a_plots}.
 The orbital angular momentum  is
 shown in the middle right  panel and we see that the orbital angular momentum is removed
 by the initial driving and falls further due to mass transfer and direct impact accretion.
 Correspondingly, the accretor is spun up, as shown in the
 bottom left panel of Figure \ref{q04a_plots}.
 The donor does not significantly change its spin angular momentum during the evolution
 as shown in the bottom right panel of Figure \ref{q04a_plots}.
  Interestingly, the accretor's spin appears to saturate
 at about 20 orbits, despite the fact that mass transfer continues.  We will return to this
 point in \S 4.

 To summarize, mass transfer grows rapidly by approximately an order of magnitude in the
 initial ten orbits of the evolution.  During this initial phase, the simulation largely follows the
 expectations for direct impact accretion: the binary loses angular momentum from the orbit which
 spins up the accreting star and the orbital separation begins to expand once the mass transfer has
 grown sufficiently (see Equation (\ref{oae}) and discussion in GPF).  Initially, $q>q_{\rm crit}\sim 0.2$
 because of the effects of direct impact (\citet{Maet04}; GPF), so the mass transfer is unstable and
 very little, if any, of the angular momentum of the stream is returned
 to the orbit. However, at approximately ten orbits,
 the mass transfer rate begins to decline and eventually the accretor's spin
 stabilizes while the binary continues to expand.   At the time that the mass transfer rate begins to
 decline, the binary mass ratio has fallen to $q = 0.396$ (from it's initial value of 0.409 as listed in
 Table 1).  This demonstrates that, for this model, the mass transfer has stabilized.
 This in turn implies that
 the critical mass ratio for stability against dynamical mass transfer must
 have grown to a value $q_{\rm crit}\ge 0.396$, which is
 the value of $q$ at 10 orbits. Although there is no driving at this stage of the evolution,
 it is possible, due to the flow induced in the donor star
 towards the
 $L1$ point, that mass transfer may continue even after $q < q_{\rm crit}$.
 The above considerations suggest that $q_{\rm crit}$ evolves and increases once the accretor spin
 saturates and a significant part of the angular momentum carried by the stream is returned to the orbit.
 We can not, based on this simulation, exclude the possibility that stability returns to the 
 binary at an even earlier point in its evolution.  Due to finite numerical resolution we can 
 only treat semi-detached systems with a mass transfer rate above a certain floor value and 
 thus the entire evolution of a real  DWD binary undergoing unstable mass transfer at rates 
 below our resolution limit is inaccessible.  It is in this sense that our simulation 
 demonstrates a bound on the fate of the system; the return to stability with the provision 
 that the mass transfer grows to amplitudes such as those shown in Figure 2.
 In the next section,
 we perform additional numerical experiments to directly measure the value of $q_{\rm crit}$
 to improve upon the lower bound we have inferred
 from the fact that the mass transfer rate declines.  The lower
 bound is still of interest though, as it is significantly higher than the value expected for negligible
 tidal coupling during direct impact accretion of $q_{\rm crit} \sim 0.2$ (\citet{Maet04}; GPF).

 \section{Driven Evolutions of the Q0.4 Binary}

We now attempt to directly measure the effective value of the critical mass ratio for stability
against dynamical mass transfer, $q_{\rm crit}$, by imposing an artificial driving loss of angular momentum.
If the model binary attains its equilibrium mass transfer rate, that rate should be simply related to measurable
quantities from the simulation and the unknown value $q_{\rm crit}$ through (see GPF)
\begin{equation}
   \dot{M}_{\rm eq} = M_{D} \frac{\left(\dot{J}/J^{\rm tot} \right)_{\rm driving}}
                                                 { \left( q_{\rm crit} - q \right) }.
   \label{meq}
\end{equation}

At a time of 7.7 orbits into the evolution, we impose one of three constant driving rates
as listed in Table 2. This time was chosen because it corresponds
to the deepest degree of contact attained during the baseline simulation before the
rate of increase of the mass transfer begins to slow down.
The baseline evolution from the previous section, where no additional driving is imposed,
is included in this section as simulation Q0.4A.
The additional
driving  is imposed in the remaining three simulations while the mass transfer rate is
still growing in the baseline calculation.  Consider the Q0.4B evolution with a driving rate
of $1.0 \times 10^{-3}$.  If $q_{\rm crit}$ followed from the limit of effective tidal coupling, we
would expect an equilibrium mass transfer rate at ten orbits of approximately $0.4\%$
of the donor's initial mass per orbit.  However, if $q_{\rm crit} = 0.4$, for example, we would
instead find a very high equilibrium mass transfer rate at ten orbits of $\sim 10\%$ of the
donor's initial mass per orbit.  Recall that from our results in the previous section, we know
that $q_{\rm crit}$ becomes $\geq 0.396$.
The driving rates of angular momentum loss listed in Table 2 are constant (independent of 
the binary or stellar parameters) and far exceed the plausible magnitude for gravitational 
radiation or magnetic braking in real binary systems.  As we have a limited range of reliable 
mass transfer rates that can be simulated, so too does this imply a range of effective driving 
rates available to manipulate the evolution of the binary.  The expression for the equilibrium 
mass transfer rate in terms of the effective value of $q_{crit}$ in Eq. (3) does not require 
the time scales for mass transfer or driving correspond with known physical mechanisms such 
as the emission of gravitational radiation.  It is merely sufficient for the mass transfer 
to have reached its equilibrium value given the applied driving loss of angular momentum.

\clearpage
\begin{table}
\caption{Driving rates for $t>7.7P_0${\tablenotemark{\dag}}}
\begin{center}
   \begin{tabular}{c|c}
      \hline
      \hline
      Simulation & $\left( \dot{J}/J^{\rm tot} \right)_{\rm driving}$ \\
      \hline
      A (Baseline) & 0.0 \\
      B & $-1.0 \times 10^{-3}$ \\
      C & $-2.0 \times 10^{-3}$ \\
      D & $-5.0 \times 10^{-3}$ \\
      \hline
   \end{tabular}
   \end{center}
   \tablenotetext{\dag}{The driving rate is given in units of $P_0^{-1}$.}
\end{table}
\clearpage

\begin{figure}[!t]
   \begin{center}
      \begin{tabular}{cccc}
         \includegraphics[scale=0.65]{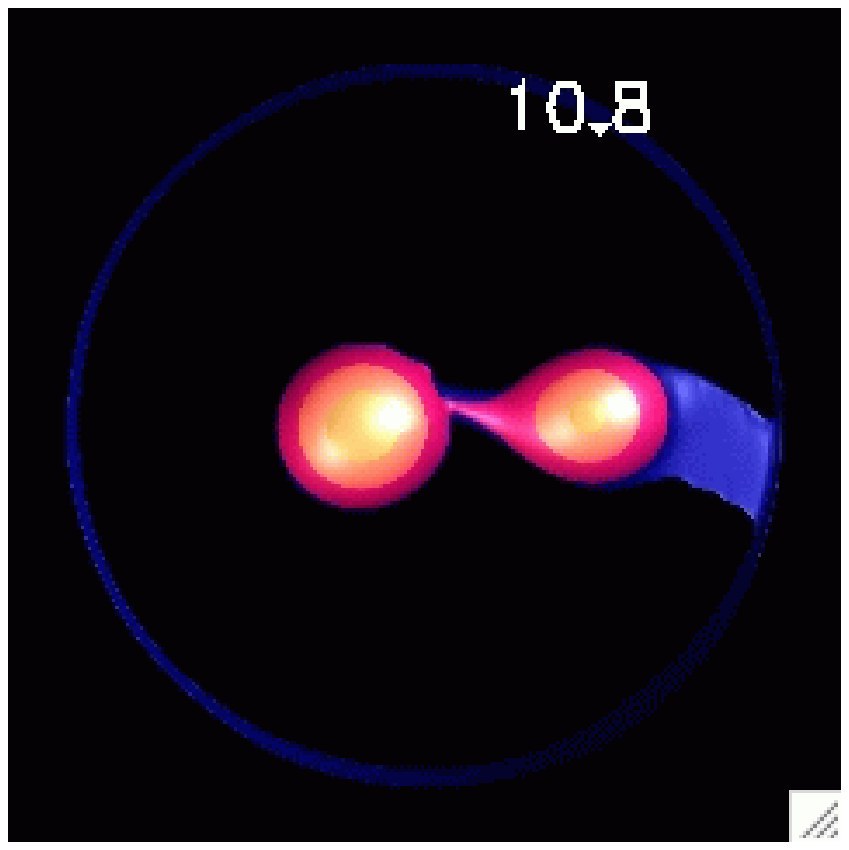} &
         \includegraphics[scale=0.65]{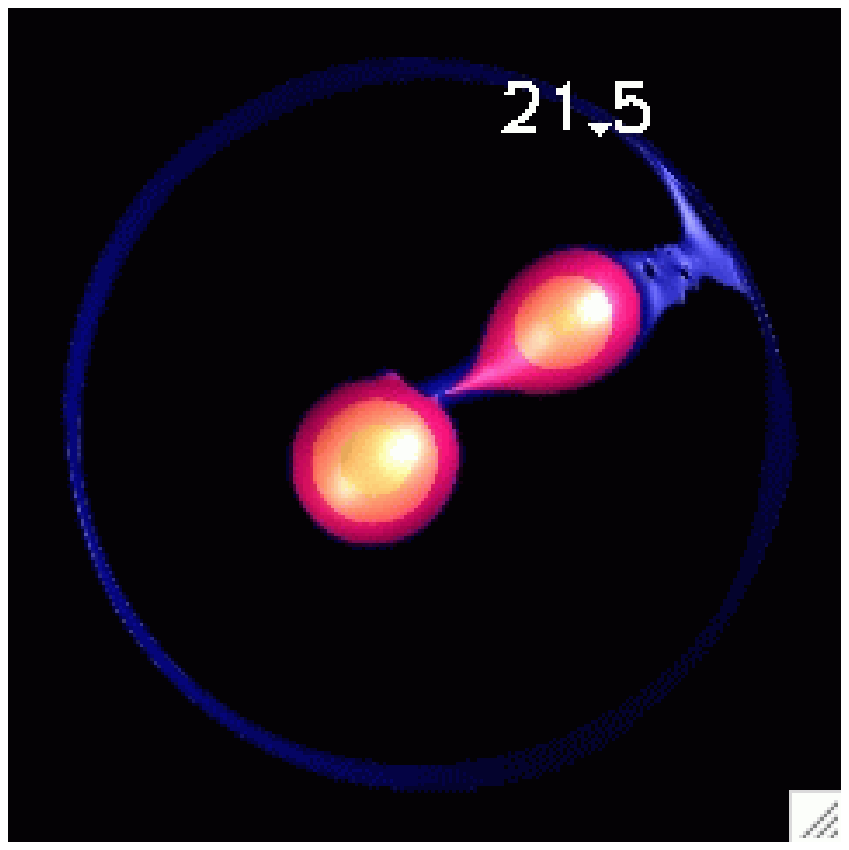} &
         \includegraphics[scale=0.65]{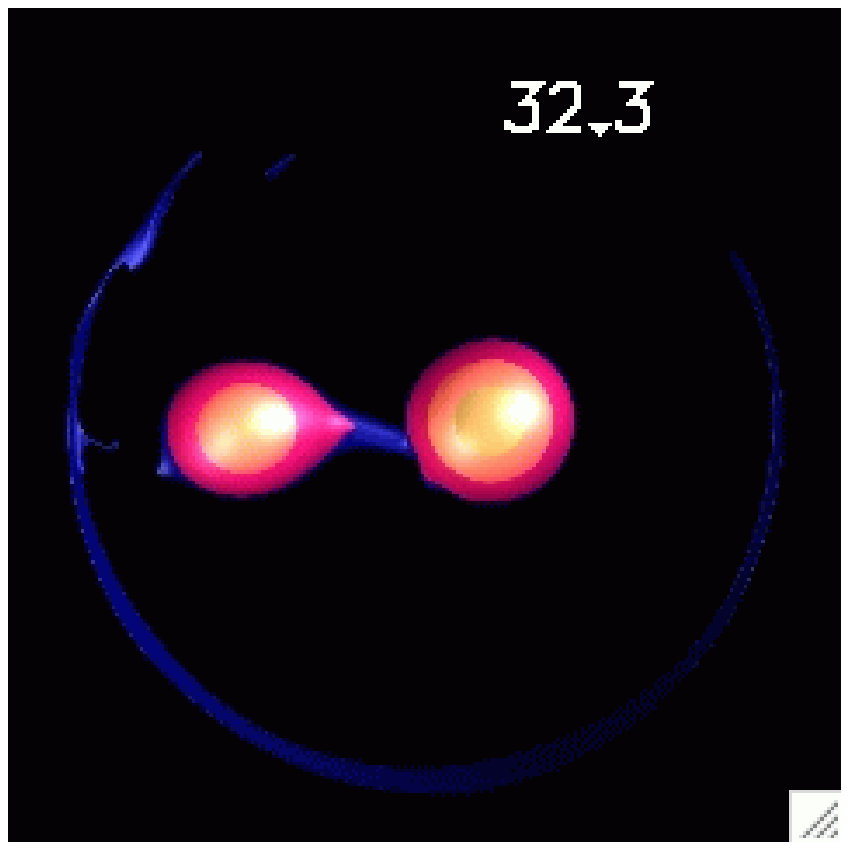} &
         \includegraphics[scale=0.65]{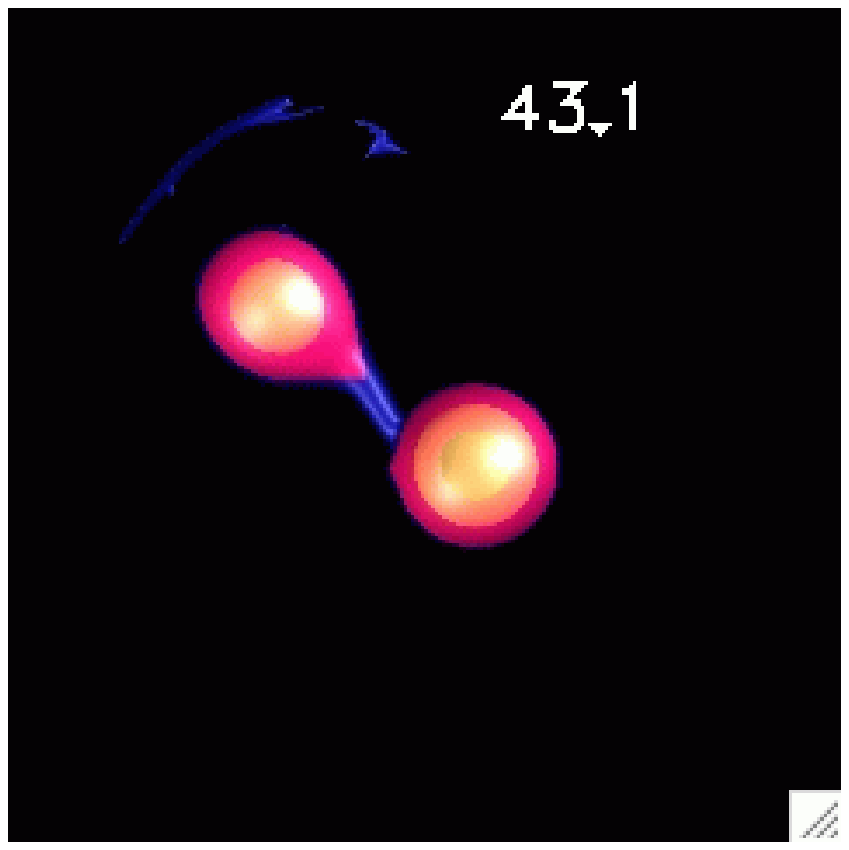}  \\
         \includegraphics[scale=0.65]{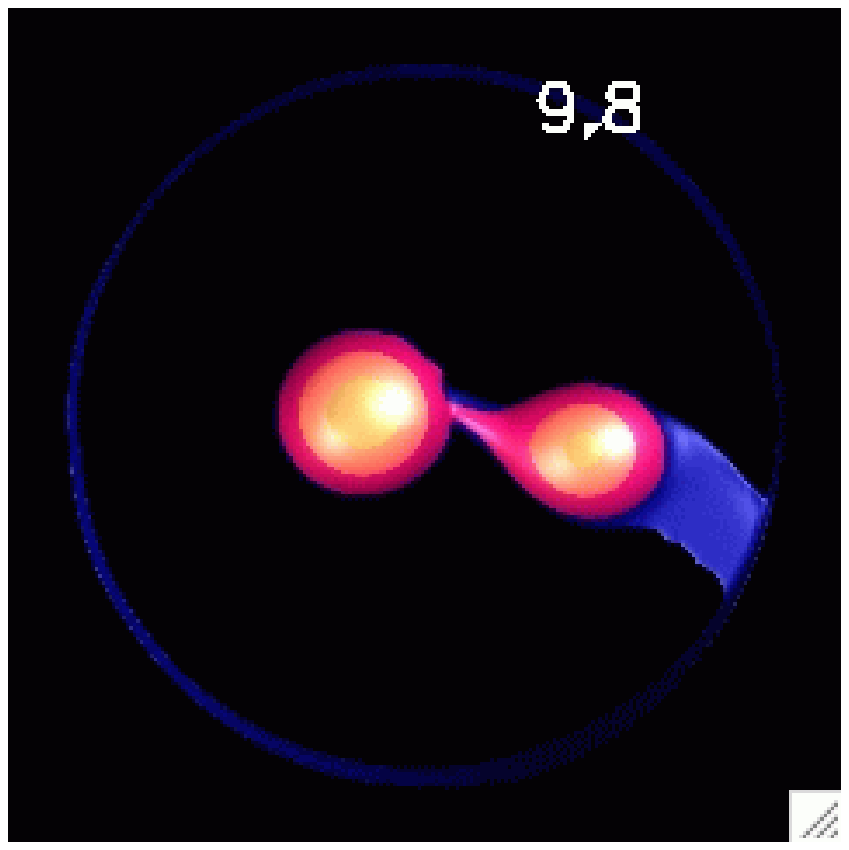} &
         \includegraphics[scale=0.65]{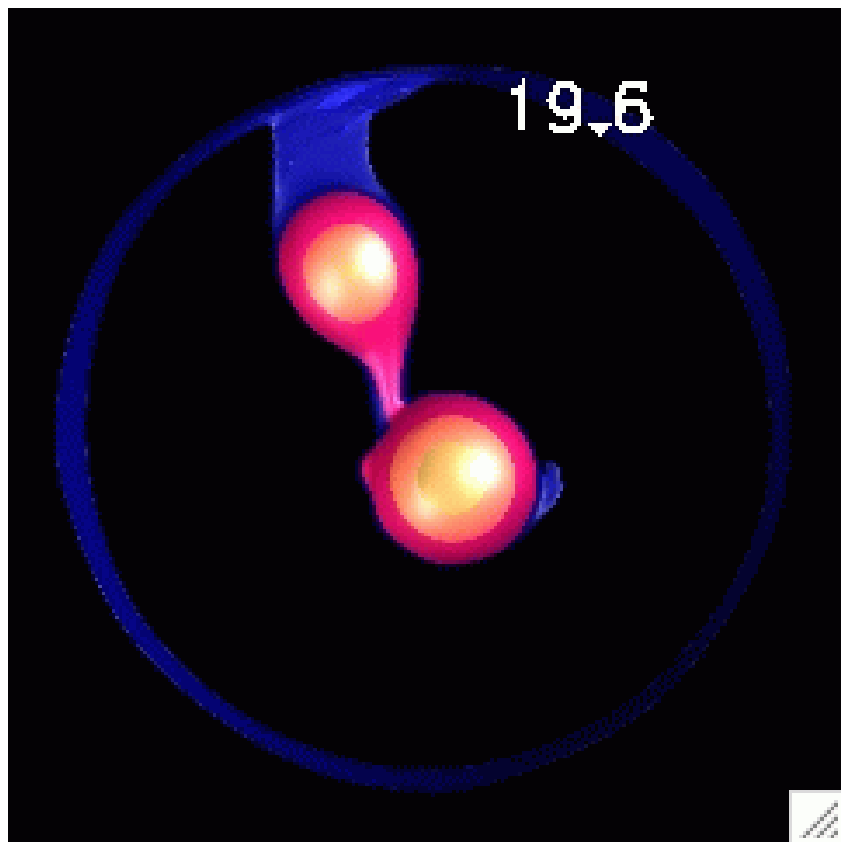} &
         \includegraphics[scale=0.65]{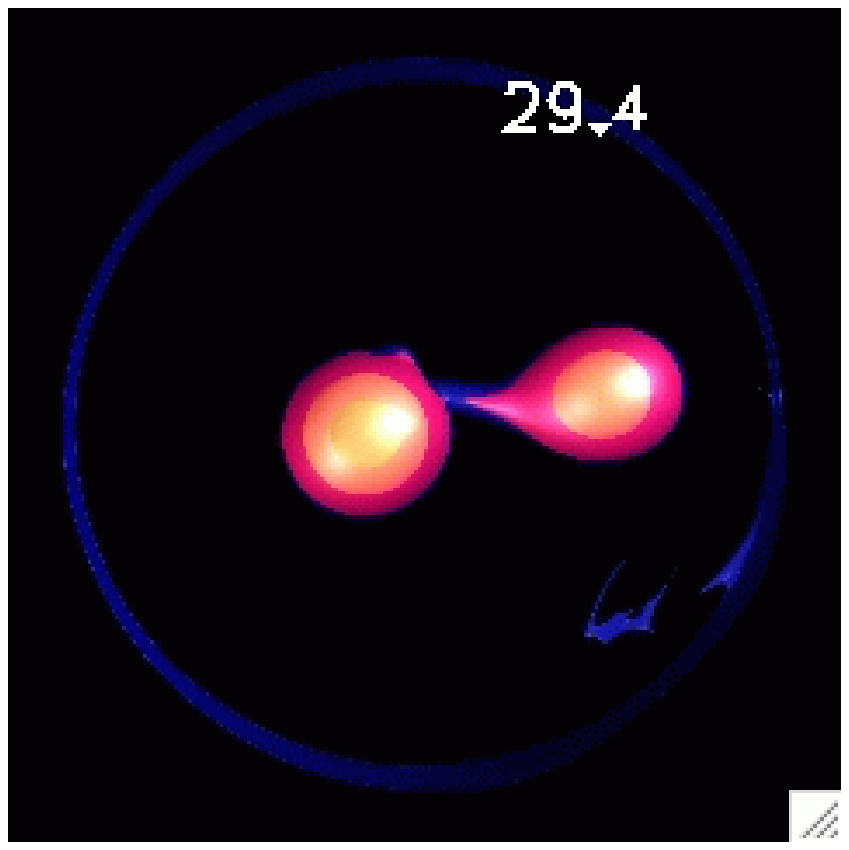} &
         \includegraphics[scale=0.65]{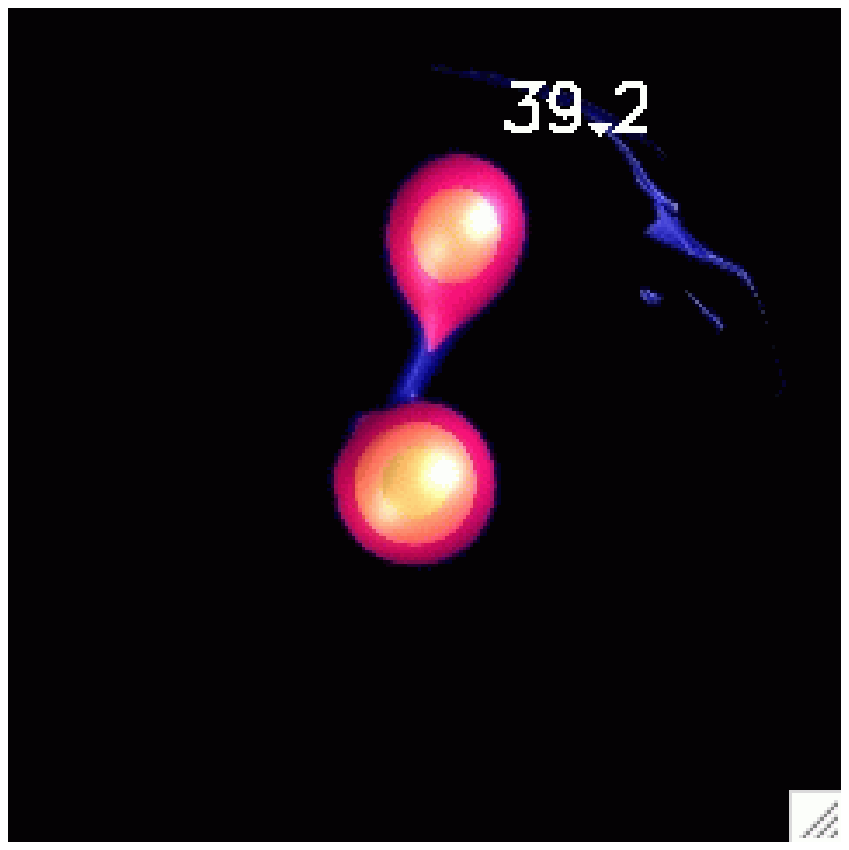}  \\
         \includegraphics[scale=0.65]{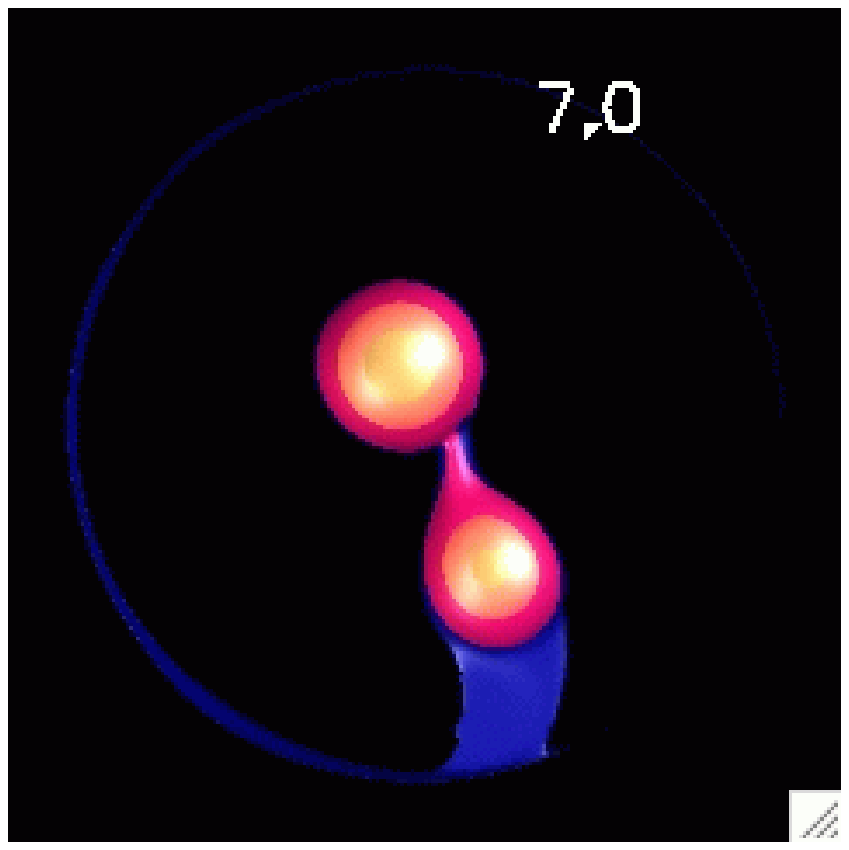} &
         \includegraphics[scale=0.65]{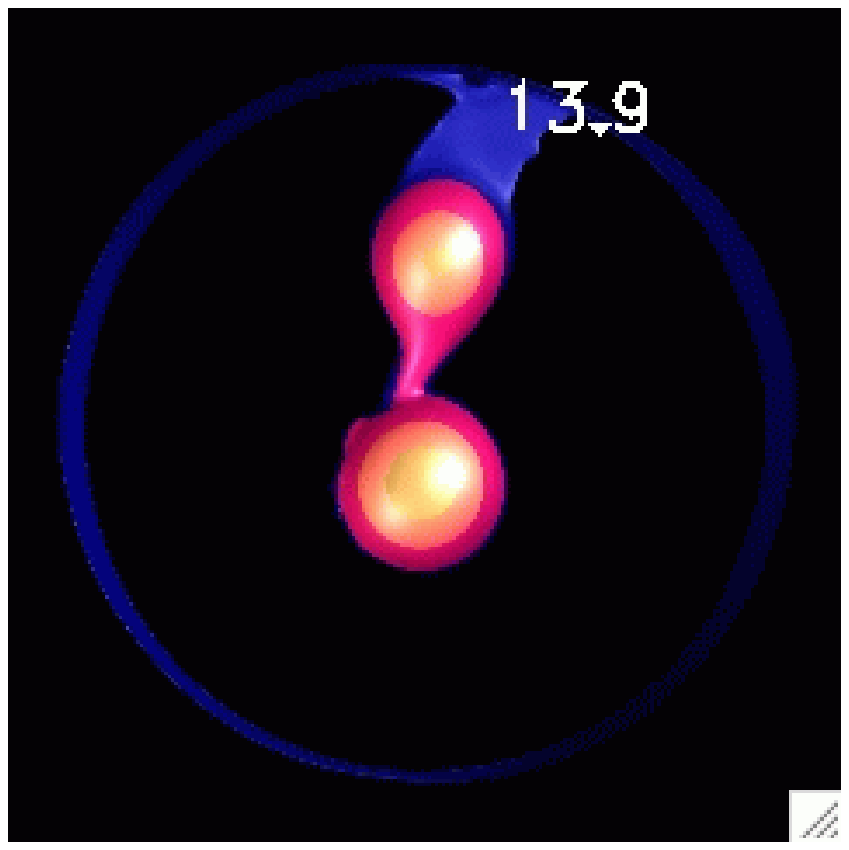} &
         \includegraphics[scale=0.65]{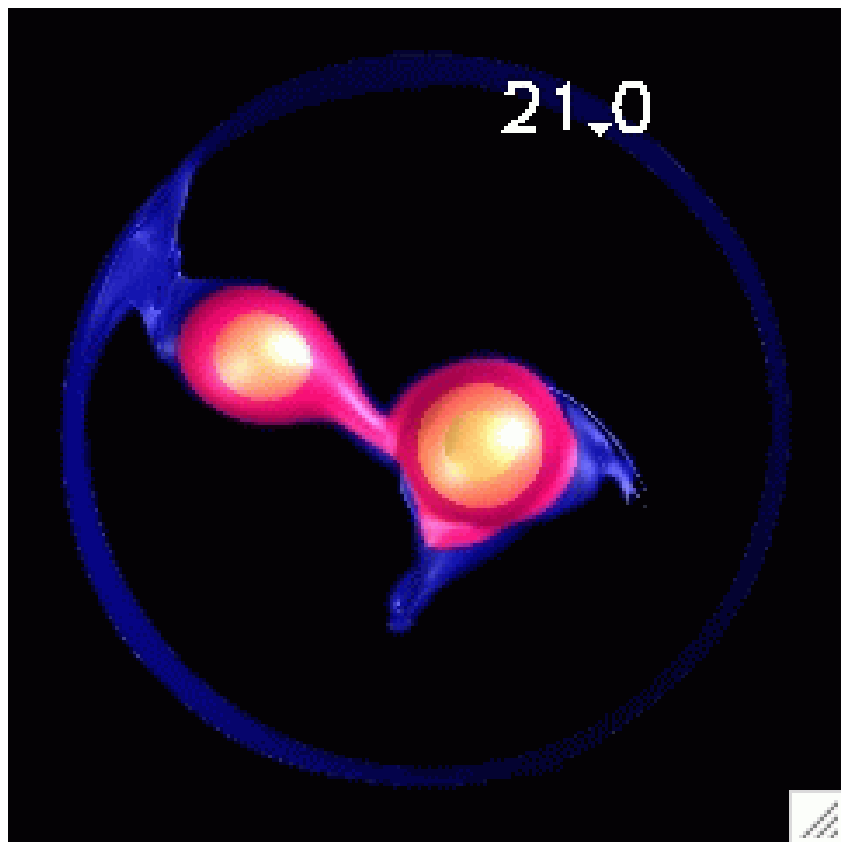} &
         \includegraphics[scale=0.65]{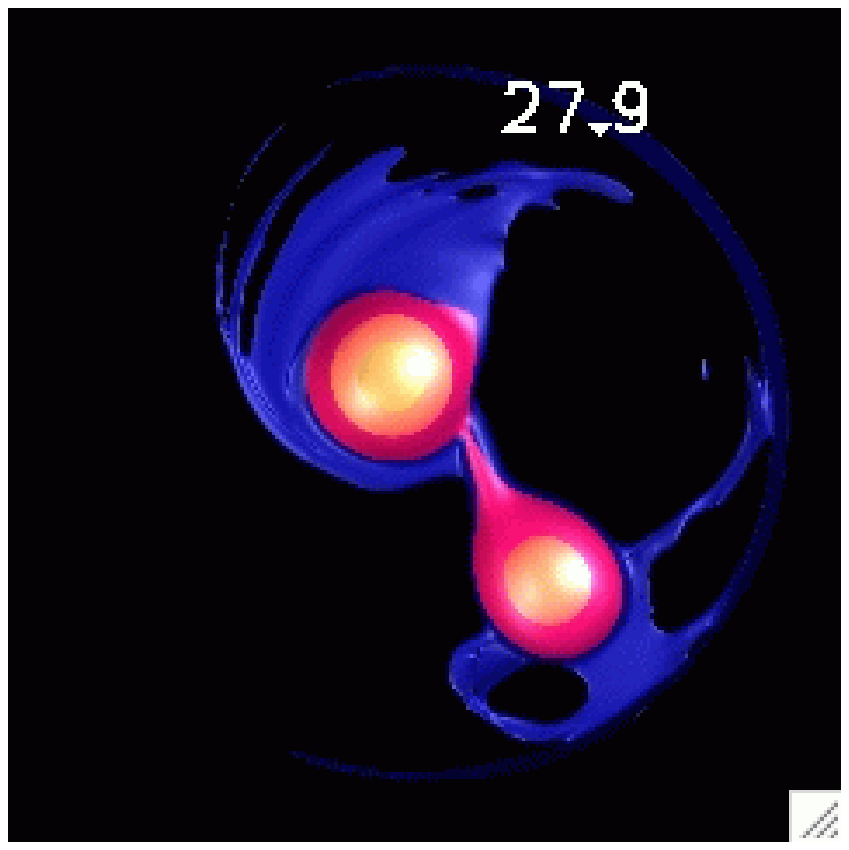}  \\
         \includegraphics[scale=0.65]{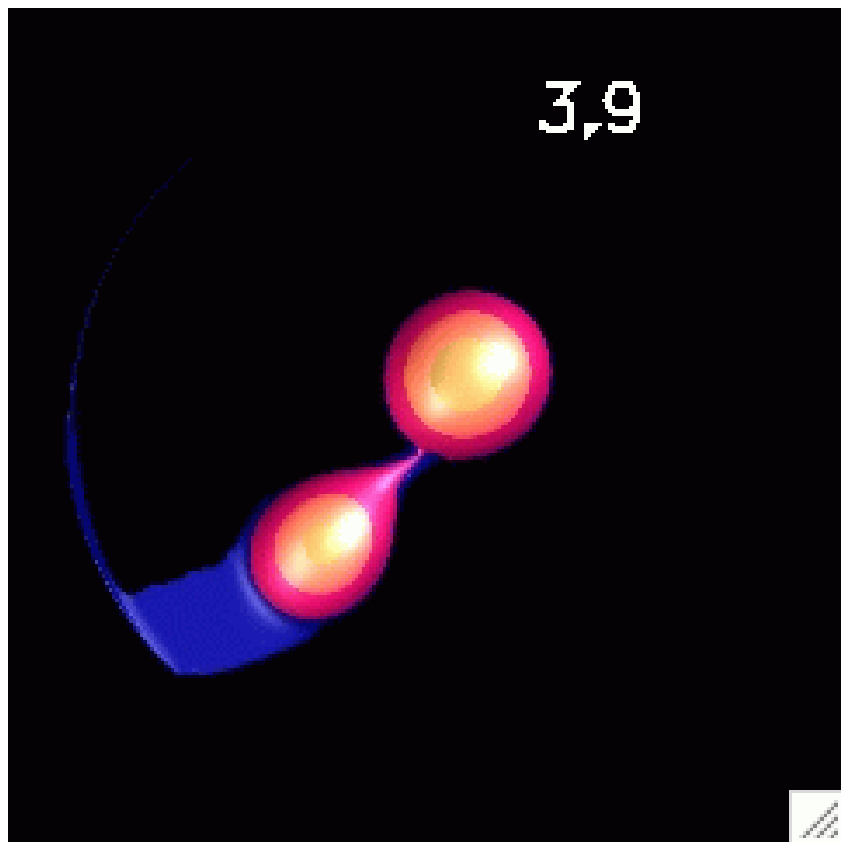} &
         \includegraphics[scale=0.65]{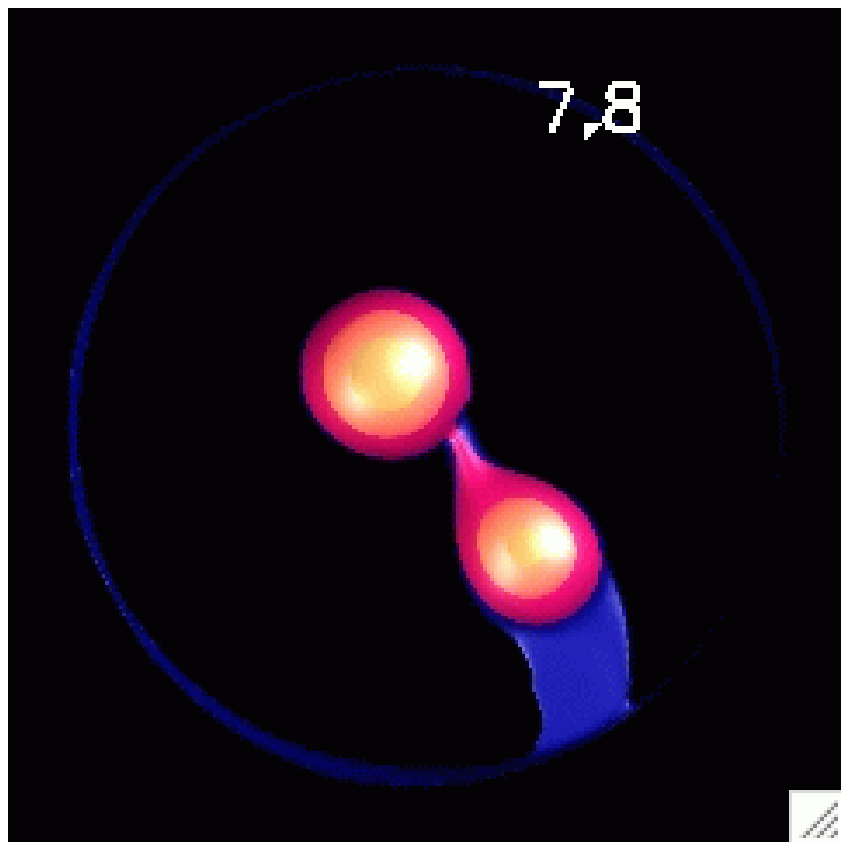} &
         \includegraphics[scale=0.65]{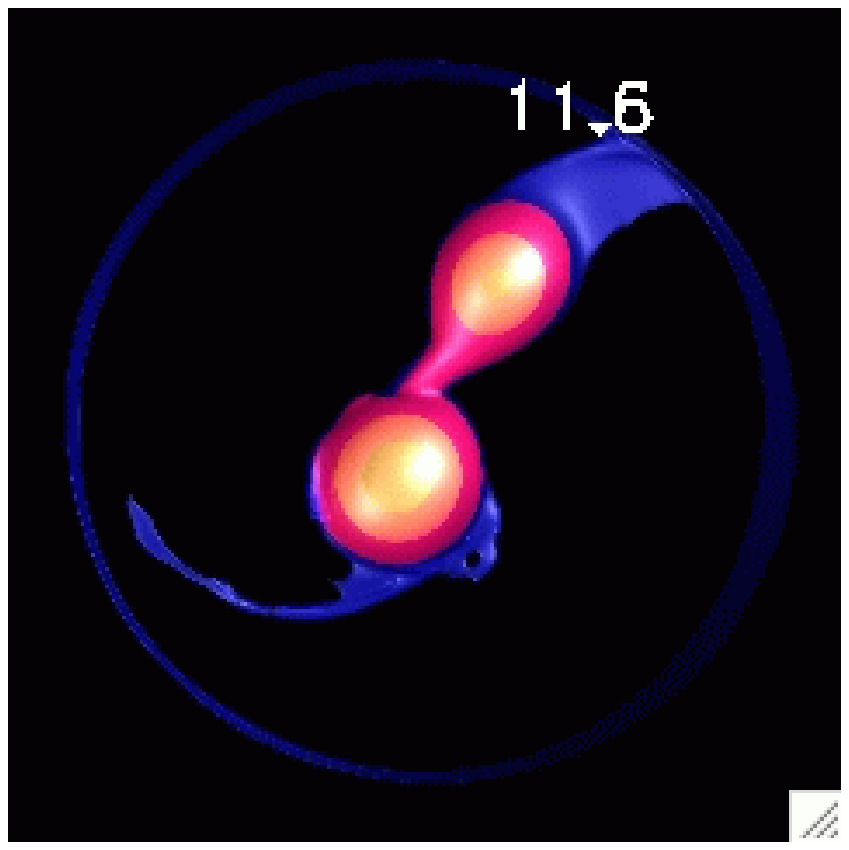} &
         \includegraphics[scale=0.65]{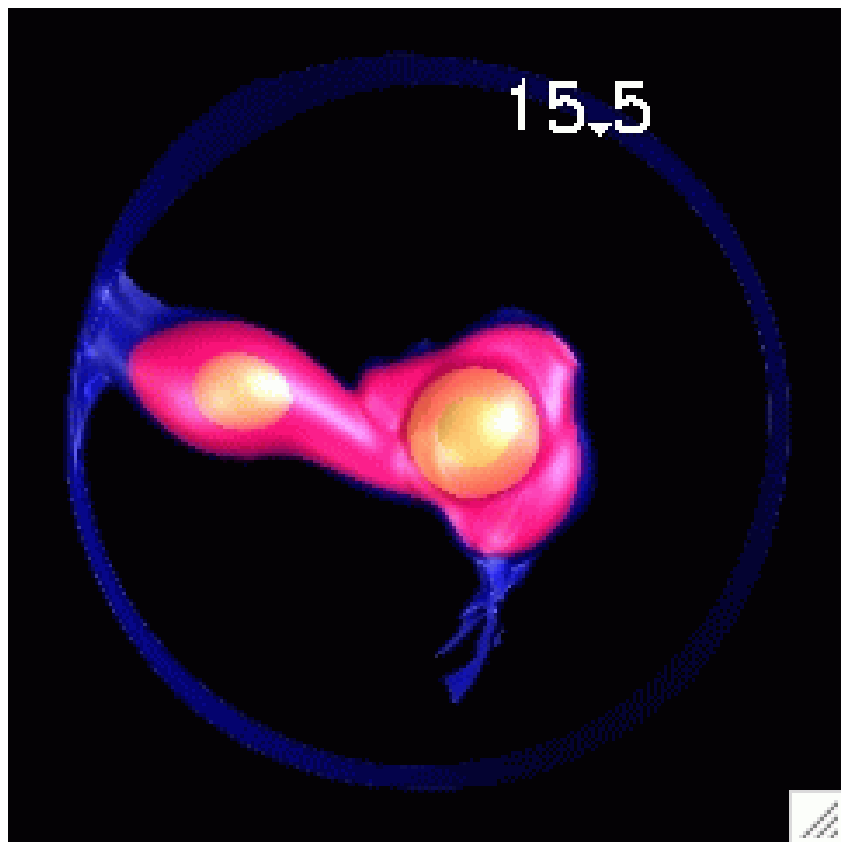}  \\
      \end{tabular}
   \end{center}
   \caption{The four rows, from top to bottom, show snapshots taken during the four evolutions Q0.4A, Q0.4B,
   Q0.4C, and Q0.4D. The snapshots show the same isopycnic surfaces as in Fig. 1, viewed from above the orbital plane,
   at approximately, from left to right, $T_{\rm f}/4$, $T_{\rm f}/2$, $3T_{\rm f}/4$, and $T_{\rm f}$,
   where $T_{\rm f}$ identifies the time at which each evolution was terminated. Each frame is labeled with
   the corresponding time in units of $P_0$. In the online
   edition, we present MPEG movies showing the complete evolutions.}
   \label{q04b_movie}
\end{figure}

Visualizations of the three driven evolutions are shown in
Figure \ref{q04b_movie}.   The images depict
the density distribution at each quarter of the respective evolution and show that as the
artificial driving rate is increased, the mass transfer becomes progressively more extreme.
The Q0.4B and Q0.4C evolutions follow the general pattern of the Q0.4A evolution in that the
mass transfer rate grows rapidly yet the binary survives to expand to larger separations
and the mass transfer event subsides.  However, in the Q0.4C run, the accretor becomes
highly distorted due to the accreted mass.
In the final simulation, at the highest driving rate
of $5 \times 10^{-3}$, the mass transfer grows to a very high level and the donor
is torn apart by tidal forces before it can escape to larger separations.

\clearpage
\clearpage

\begin{figure}
   \begin{center}
      \begin{tabular}{cc}
         \includegraphics[scale=0.4]{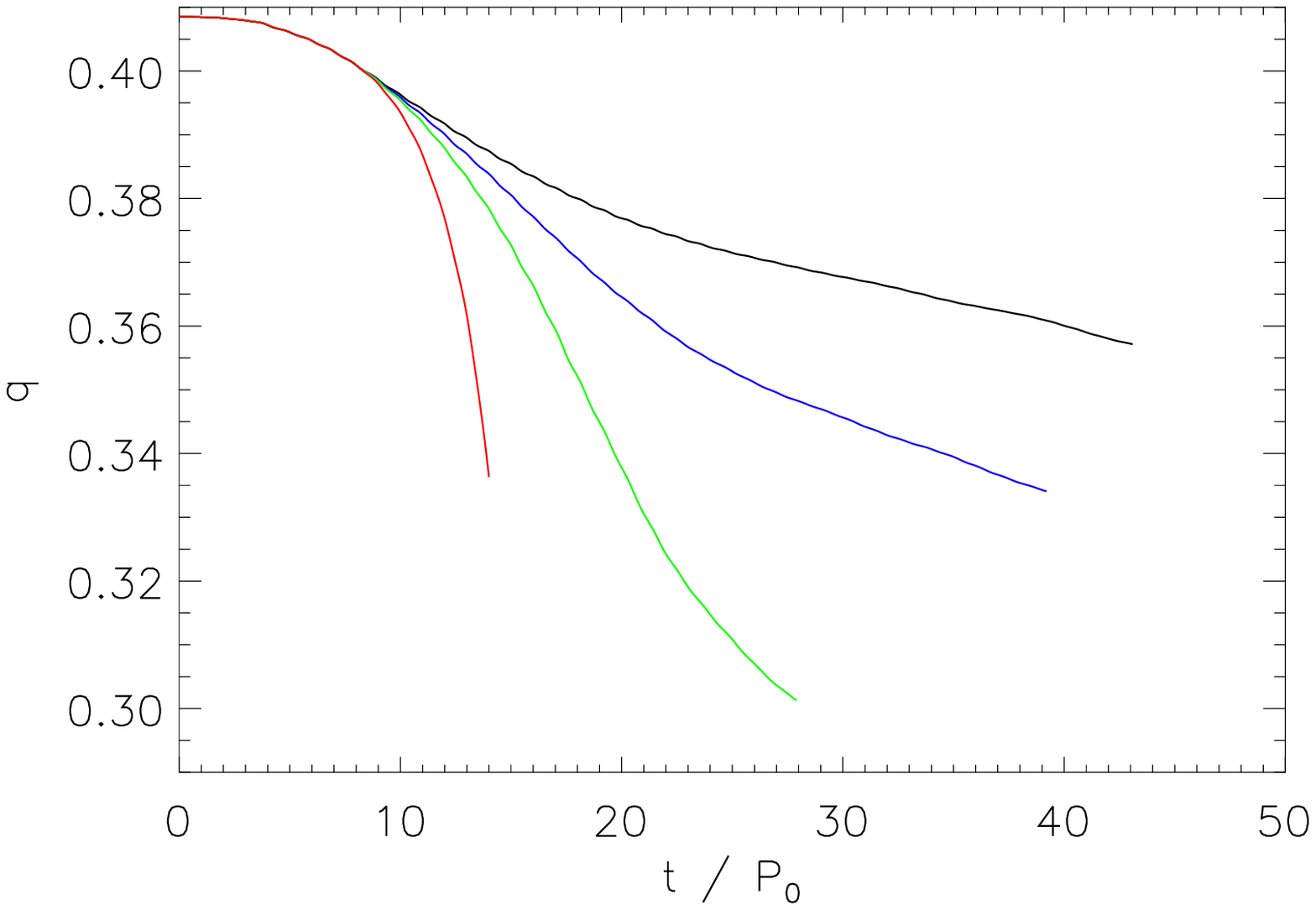} &
         \includegraphics[scale=0.4]{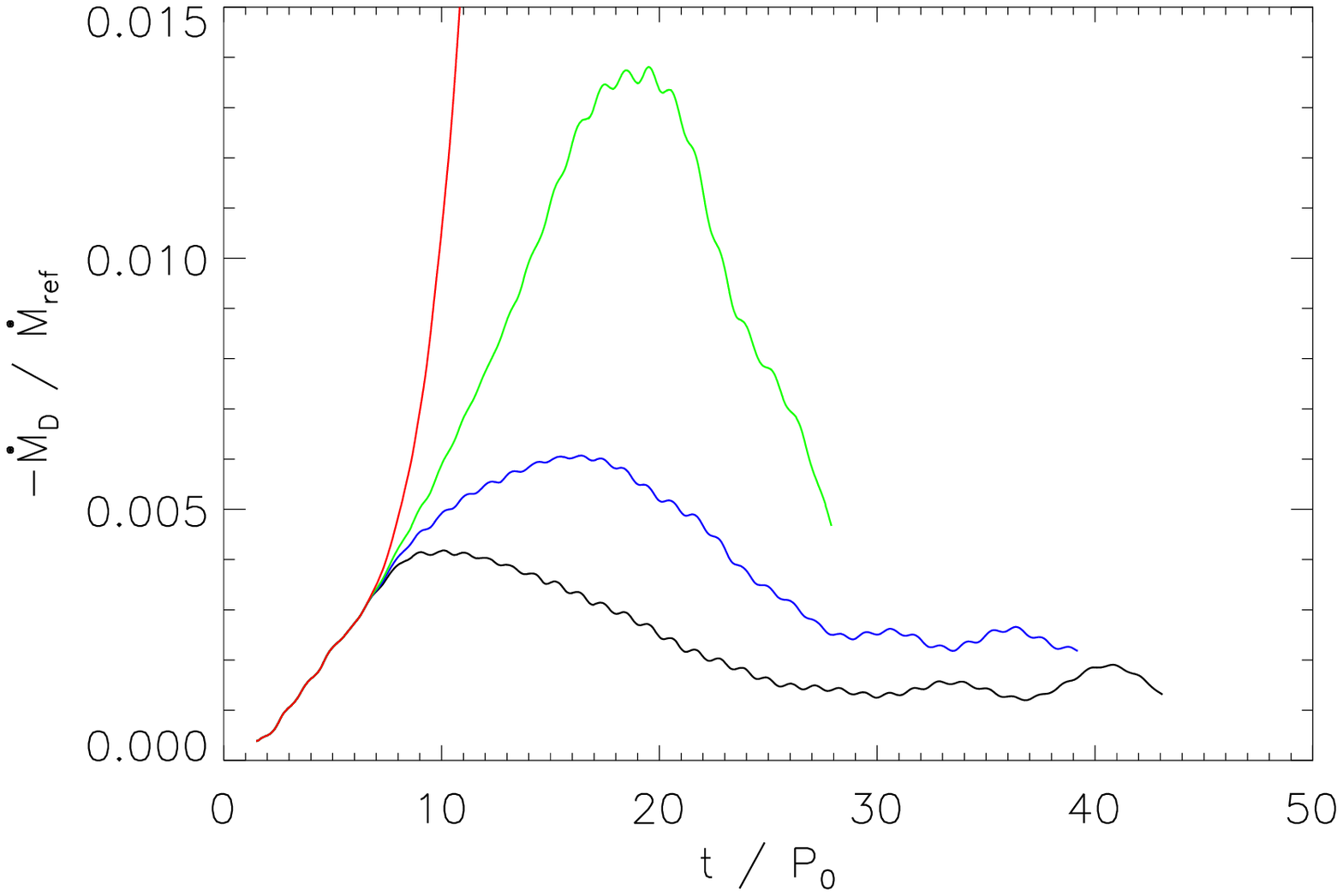} \\
         \includegraphics[scale=0.4]{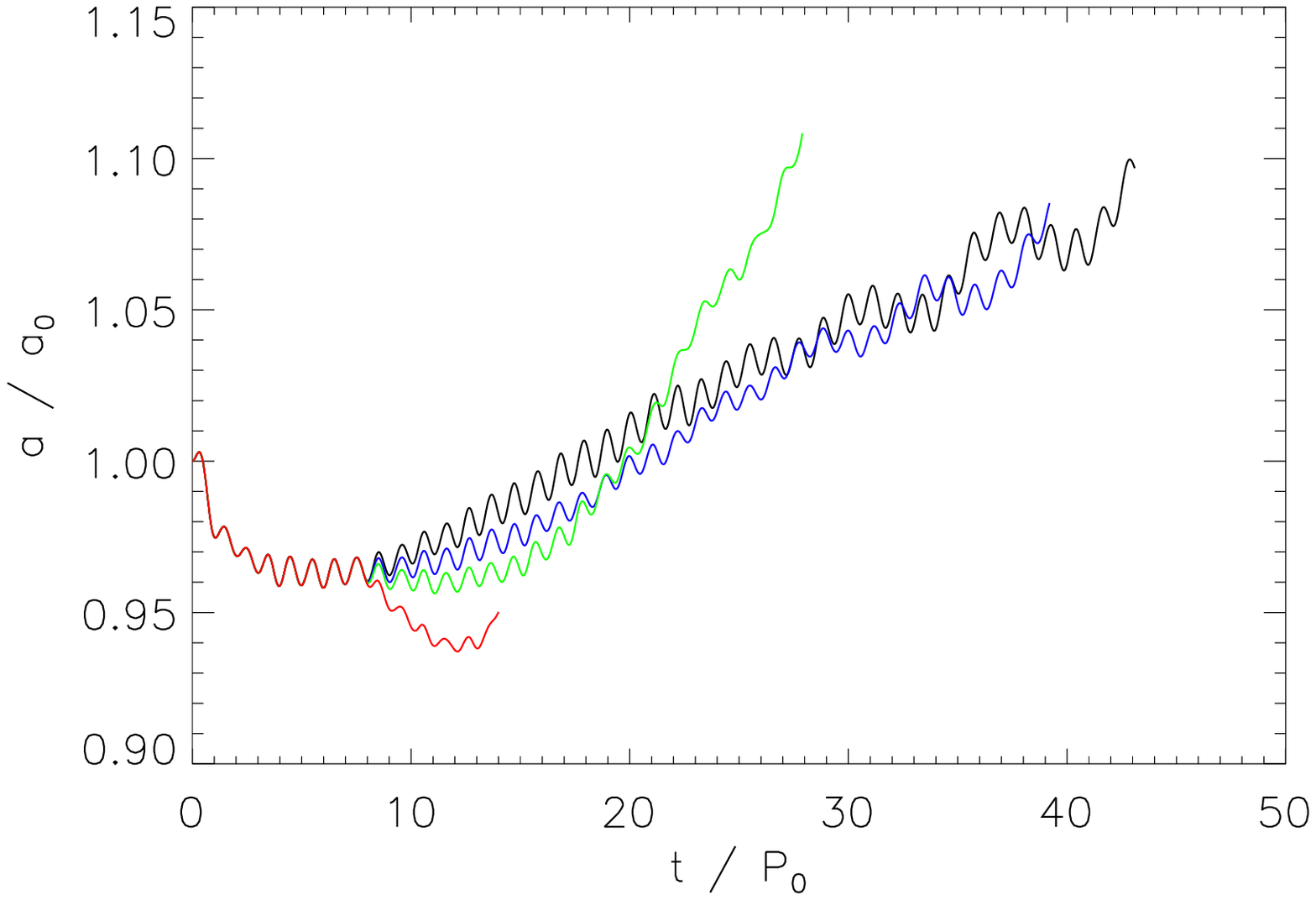} &
         \includegraphics[scale=0.4]{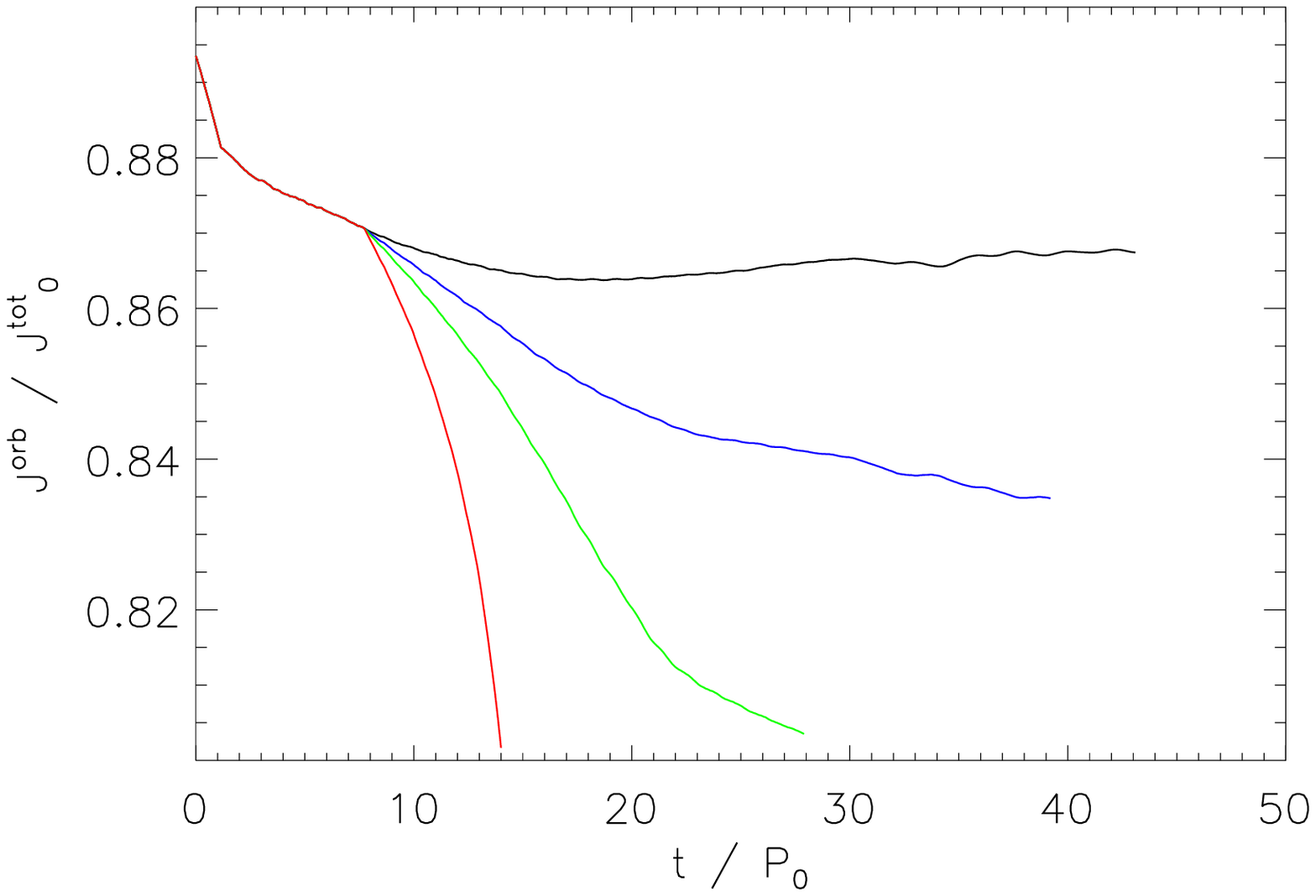} \\
         \includegraphics[scale=0.4]{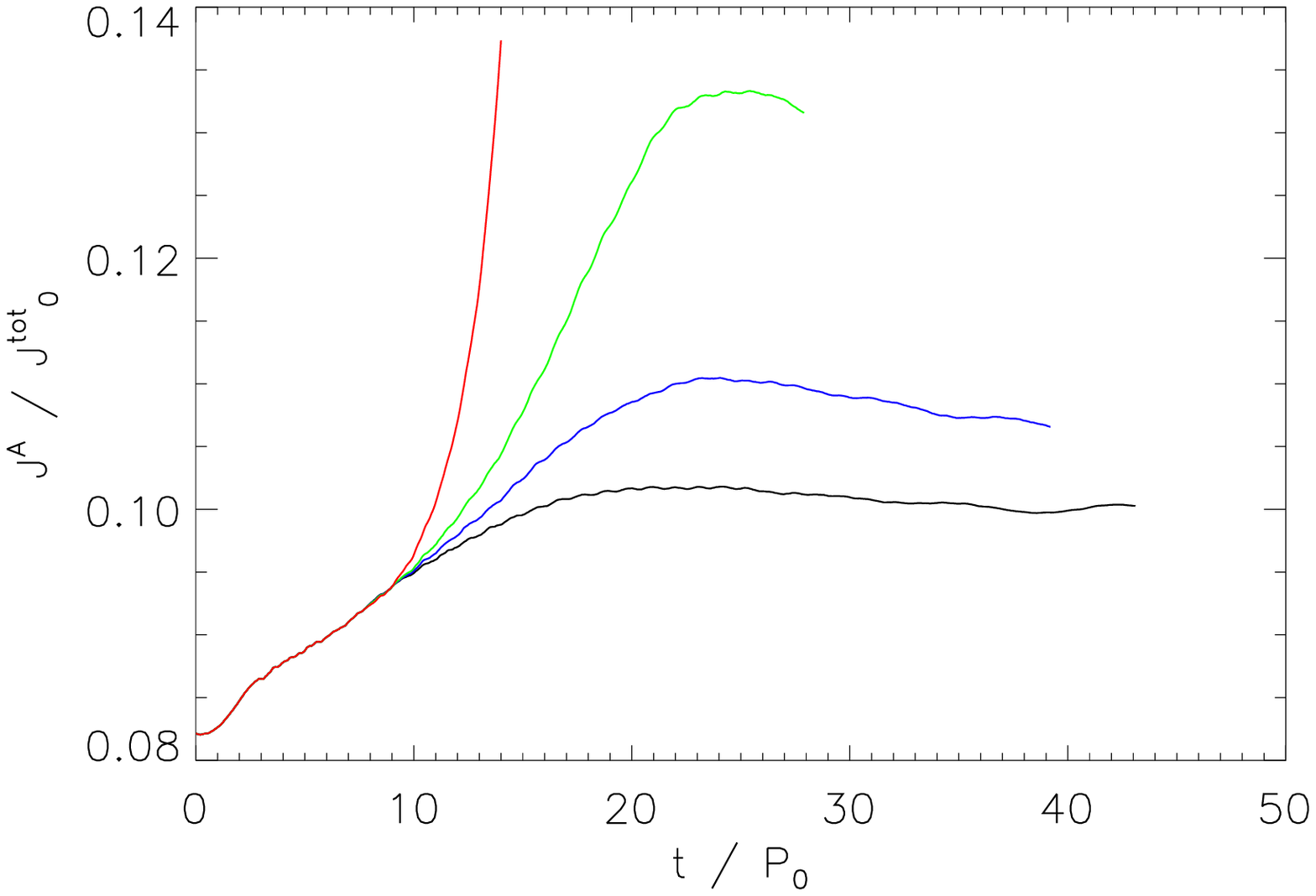} &
         \includegraphics[scale=0.4]{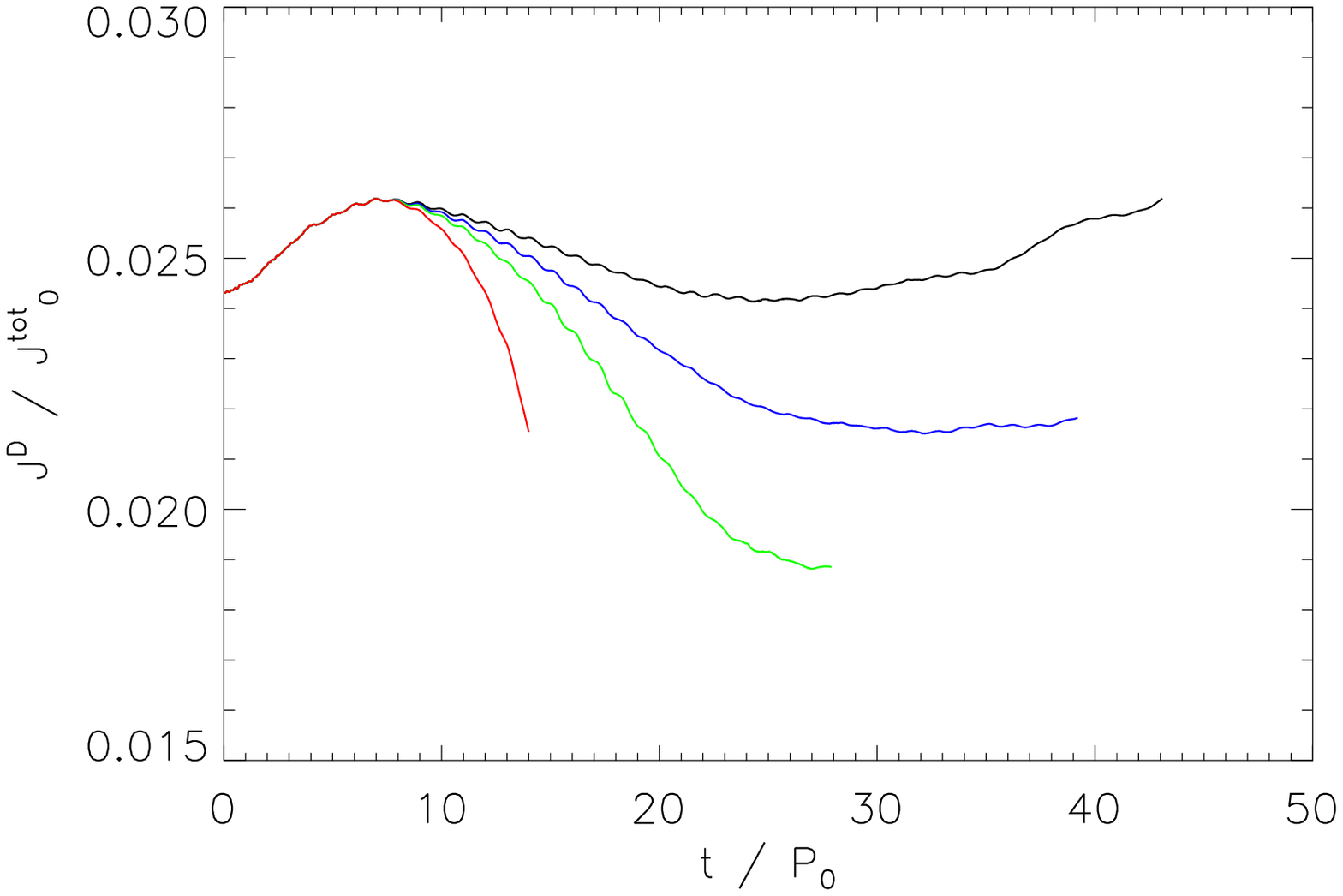} \\
      \end{tabular}
   \end{center}
   \caption{Binary parameters, as plotted in Figure 2, for the baseline evolution Q0.4A (black
   curves) and the three driven evolutions Q0.4B (blue curves), Q0.4C (green curves) and
   at the highest driving rate considered for the Q0.4D evolution (red curves).  }
   \label{ensem_plots}
\end{figure}

As shown in Figure \ref{ensem_plots},
the trajectories of the key binary parameters confirm
the impression that is drawn from a visual inspection of the images of the individual runs.
(The key parameters displayed in the six panels of Figure \ref{ensem_plots} are the same as in Figure 2
and the trajectories for the baseline evolution Q0.4A, shown in Figure 2, have been re-drawn as black curves
for comparison.)
The behavior of model Q0.4D, that was driven at the largest rate,
is plotted as the red curves in the six panels of
Figure \ref{ensem_plots}.  The binary mass ratio plunges catastrophically as the advected
angular momentum carried by the accretion stream spins up the accretor.  In this simulation, the
binary separation falls and then tries to expand but, by this point, the donor star has insufficient
mass to bind the star and the donor is shredded by tidal forces.  At the two lower driving
rates we have considered, the binary is able to remain intact.  The green curves correspond
to the Q0.4C binary which was driven at $0.2\%$ from 7.7 orbits onwards.  The high mass
transfer rate causes the binary to rapidly separate and even in this case, the accretor's spin
angular momentum stalls.  While this evolution is of note because the binary survives, the
system does not attain steady state mass transfer and rapidly encroaches on the grid
boundary, forcing us to end the evolution.

Results from the binary driven at the lowest rate, the Q0.4B binary which was driven at $0.1\%$
from 7.7 orbits onwards, are shown as the blue curves in Figure \ref{ensem_plots}.  This evolution
closely follows the results we have presented in \S2 for the baseline evolution:
The mass transfer rate rises initially, then falls to an approximately steady level as the
binary  expands.
At the same time, the binary loses orbital angular momentum into the spin of the accreting star
which again appears to saturate and become approximately constant.

If we take the mass transfer rate in the Q0.4B binary from approximately 30 orbits onwards as
the equilibrium mass transfer rate, $\dot{M}_{\rm eq}$, we see immediately that the low
mass transfer
rate in this phase is not consistent with a small value for $q_{\rm crit} \sim 0.4$.   Taking the
driving rate at $10^{-3}$ and using values for  the mass ratio and mass transfer rate from
the plots in Figure \ref{ensem_plots} in Equation (\ref{meq}), we deduce that
$q_{\rm crit} \sim 0.7$.  This value is consistent with the expectation from Equation (\ref{qdisk})
for very efficient return of angular momentum as would be the case when an accretion
disk is present that fills most of the accretor's Roche lobe. No such disk develops in
evolution Q0.4B, however.

\clearpage
\clearpage

For this model binary, with an initial mass ratio intermediate between the plausible maximum
and minimum values of $q_{\rm crit}$, we find that it behaves, in the steady phase of the
evolution,  in a manner  more
consistent with that case of maximum efficiency for tidal coupling.
The advected angular momentum that spins
up the accretor is returned to the binary orbit on a timescale comparable to the mass
transfer timescale itself.  We therefore conclude that in many instances, DWD binaries could
survive to reach a long-lived AM CVn state if hydrodynamics and gravity were the only
processes in play.  To understand this curious result in more detail,
we turn in the next section to analyze the simulation output from the point of view of a relatively
simple, orbit-averaged equation of motion for the binary.

 \section{Comparison to Expectations from the Orbit-Averaged Equations}

 In the previous  sections, we have demonstrated the surprising resiliency of a polytropic
 binary that undergoes direct impact accretion.
 We have found that, in terms of dynamics, the binary behaves as if there were 
 a mechanism in the system that can  effectively return angular momentum from the
 accretor's spin to the binary orbit.  We have also observed in the three simulations of this
 binary where the donor star survives, that the accretor does not spin up indefinitely.
 To the contrary, the spin angular momentum of the accretor appears to saturate
 despite the fact that
 mass transfer continues to bring high-angular momentum material onto the accretor.

 To understand this behavior, we turn to a simplified orbit-averaged equation to
 analyze the  different physical effects that are competing to determine the fate of
 the binary.  For general background on orbit-averaged equations and their
 application to the orbital evolution of DWD see GPF. We
 first decompose the angular momentum of the binary into the orbital angular momentum of
 each star's center of mass and the spin angular momentum of each star about their
 respective center of mass as follows
 \begin{equation}
    J^{\rm tot} = J_{\rm orb} + J_{A} + J_{D}.
    \label{j}
 \end{equation}
 For a point mass binary, we may use Kepler's third law to express the orbital angular
 momentum as
  \begin{equation}
    J_{\rm orb} = M_{A} M_{D} \sqrt{ \frac{ G \;  a }{ M_{A} + M_{D} } },
    \label{k3}
 \end{equation}
 where $a$ is the orbital separation and $G$ is the universal gravitational constant.
 Combining Equations (\ref{j}) and (\ref{k3}) and then logarithmically differentiating
 with respect to time we obtain
 \begin{equation}
    \label{oae}
    \frac{\dot{a}}{2a} =    \left( \frac{\dot{J}_{\rm orb}}{J_{\rm orb}} \right)_{\rm driving}
                                        - \left( \frac{\dot{J}_{A}}{J_{\rm orb}} + \frac{\dot{J}_{D}}{J_{\rm orb}} \right)
                                        - \frac{\dot{M}_{D}}{M_{D}} \left( 1 - q \right),
\end{equation}
where we have assumed conservative mass transfer.

The left hand side in Equation
(\ref{oae}) reflects the trajectory of the mean orbit and this is equated to several
competing terms on the right hand side.  Since $\dot{J}_{\rm orb}$ is negative, the driving term
tends to shrink the binary while the exchange of mass tries to expand the system since
the mass ratio is less than unity.  Recall that $\dot{M}_{D}$  is by its nature a negative
quantity.
On the other hand, an increase in the spin angular momentum of either
the donor or accretor drives $\dot{a}$ to negative values as the angular momentum
had to ultimately come at the expense of the binary orbit.

We apply Equation (\ref{oae}) by simply substituting the simulation data (as appears in the
plots of Figure \ref{ensem_plots}) for all the
terms.  Time derivatives are evaluated through numerical differentiation and the natural
scale of the terms is the initial binary frequency, $\nu_{0} = 1 / P_{0}$.  For example, the
term on the left hand side is proportional to the rate at which the binary separation changes
as a percentage rate per orbital period, $P_{0}$.
The data values and
their time derivatives  have been
smoothed with a sliding boxcar average with a width of three $P_{0}$,
as was used previously to construct the
mass transfer rates in Figures \ref{q04a_plots} and \ref{ensem_plots}.

\begin{figure}[!t]
   \begin{center}
       \plotone{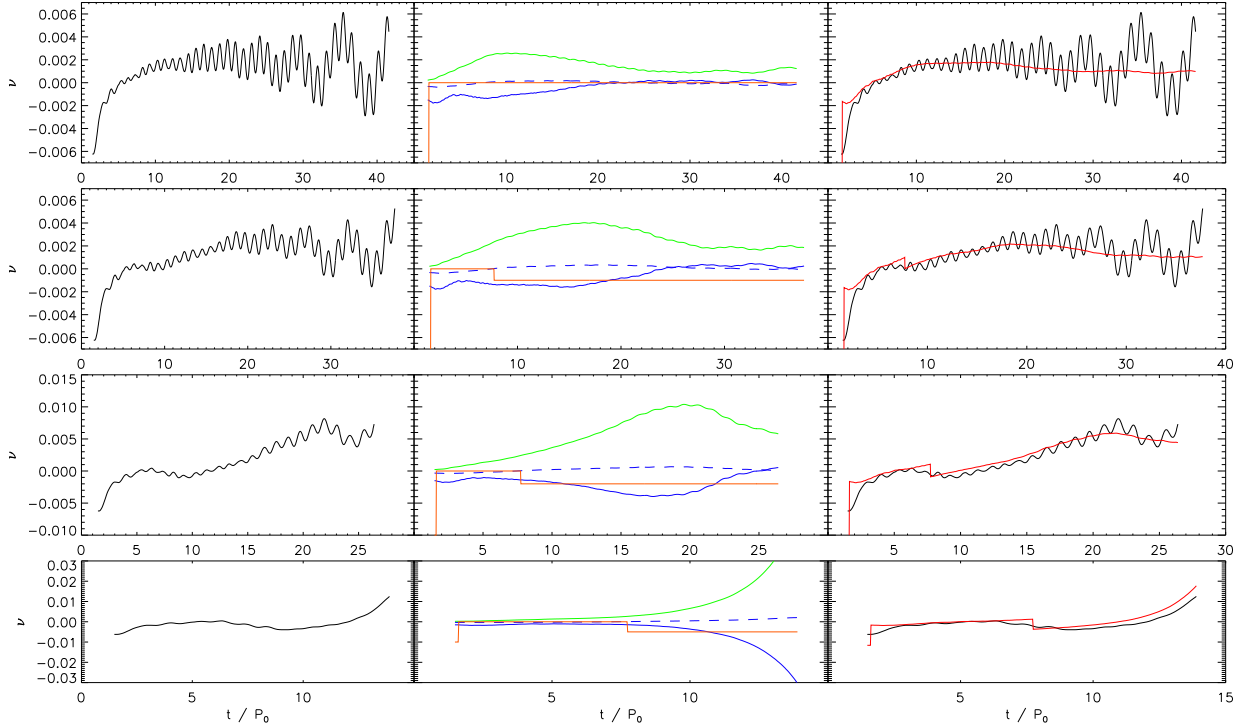}
   \end{center}
   \caption{Terms appearing in the orbit averaged Equation \ref{oae} for the four evolutions
   presented in this paper. The four rows, from top to bottom, correspond to evolutions
   Q0.4A, Q0.4B, Q0.4C, and Q0.4D.
   The left-hand side of
   Equation \ref{oae}, which traces variation in orbital separation, is shown as the oscillatory
   black curve on the left column.  The central column shows the individual terms on the right-hand side
   of the same equation: the driving angular momentum term is shown in orange and the mass transfer term is plotted in green; the
   accretor and donor spin terms are plotted as a solid and a dashed blue line respectively. The right column
   shows superimposed the left-hand side of the equation (black) and the sum of all terms on the right hand side (red),
   to demonstrate that Equation \ref{oae} is indeed satisfied on average. The terms in Equation \ref{oae} have a dimension of
   inverse time and here, time is measured in units of the initial orbital period. The ordinate $\nu$ is plotted
   in units of $P_0^{-1}$.}
   \label{oae_plots}
\end{figure}

The terms appearing in Equation (\ref{oae}) are plotted as a function of time for the four simulations
in Figure \ref{oae_plots}; each row in the figure represents a separate simulation where
Q0.4A is across the top and Q0.4D is across the bottom.
The black curves in both the left-hand panels and the right-hand panels of Figure \ref{oae_plots}
display the behavior of the left-hand side of Equation (\ref{oae}), while the total sum
of the terms on the right-hand side of that equation is shown in red on the right-hand panels of the figure.
The two sides of the equation
do indeed track each other, with the orbit oscillating (due to epicyclic motion)
about  the mean dictated by the right hand side.
The behaviors of the individual terms that contribute to the right-hand side of Equation (\ref{oae})
are displayed in the middle column of Figure \ref{oae_plots}.
The driving loss of angular momentum
appears in orange and is given by a superposition of step functions.  The initial driving to
force the binary into contact appears in all four simulations at a $1\%$ rate of loss for the
initial 1.6 orbits.  At 7.7 orbits, the additional driving terms are switched on at the rates
given in Table 2 for simulations Q0.4B - Q0.4D.

The mass transfer term, that is, the last term appearing on the right hand side of Equation
(\ref{oae}), is plotted (including the sign) as the green curves in Figure \ref{oae_plots}.
This term is indeed stabilizing and tries to force the binary separation to expand.
The spin terms are shown in blue with the donor's spin (again, including the sign)
plotted as dashed blue lines near zero amplitude and the accretor's spin contribution
plotted as solid blue curves.

\clearpage
\clearpage

 \begin{table}\caption{Reference times}\vspace{10pt}
    \begin{center}
       \begin{tabular}{cccc}
       \hline\\[-1.5ex]
       Simulation & $\left( \frac{ \dot{J}}{J} \right)_{\rm driving}$ &
       Time of Minimum $\left(-\frac{ \dot{J}_{A}}{ J_{\rm orb}}\right) $ &
       Time of Maximum $\left(-\frac{ \dot{M}_{D}}{M_{D}} \left( 1 - q \right)\right) $ \\
       \hline
       \hline
       Q0.4A & 0.0 & 7.94 $P_{0}$ & 10.12 $P_{0}$ \\
       Q0.4B & $1 \times 10^{-3}$ & 14.29 $P_{0}$ & 17.34 $P_{0}$ \\
       Q0.4C & $2 \times 10^{-3}$ & 17.37 $P_{0}$ & 19.56 $P_{0}$ \\
       \end{tabular}
    \end{center}
 \end{table}
\clearpage

 The simulated binaries satisfy Equation (\ref{oae}) to a surprising accuracy.  The
 black curves closely track the effective driving within the binary - the sum of all
 terms on the right hand side plotted in red - even for the extreme mass
 transfer of the Q0.4D simulation.  For this run, the mass transfer and accretor
 spin terms each
 diverge in opposite directions and yet the binary separation tracks their combined
 effect closely.   In the remaining three simulations, the binary separation oscillates
 about the effective driving as eccentricity builds up in the orbit.

 We also observe that while the spin-up of the accretor tries to destabilize the binary, this
 term makes its largest contribution to the balance
 of Equation (\ref{oae}) relatively early in each evolution and then begins
 to decrease to zero.  The accretor's spin term rising to zero means that the spin angular momentum
 of the accretor is approaching a constant value.  Shortly after the accretor spin term
 starts to diminish, the mass transfer term also begins to decrease in all three simulations
 where the donor star survives.  For reference, we have recorded the time of the largest
 magnitude contribution from the accretor's spin term and mass transfer term in Table 3.
 In all three cases where the donor survives, the accretor spin up begins to decrease and
 2-3 orbits later, the mass transfer rate begins to decline as well.

\begin{figure}[!t]
   \begin{center}
       \includegraphics[scale=0.6]{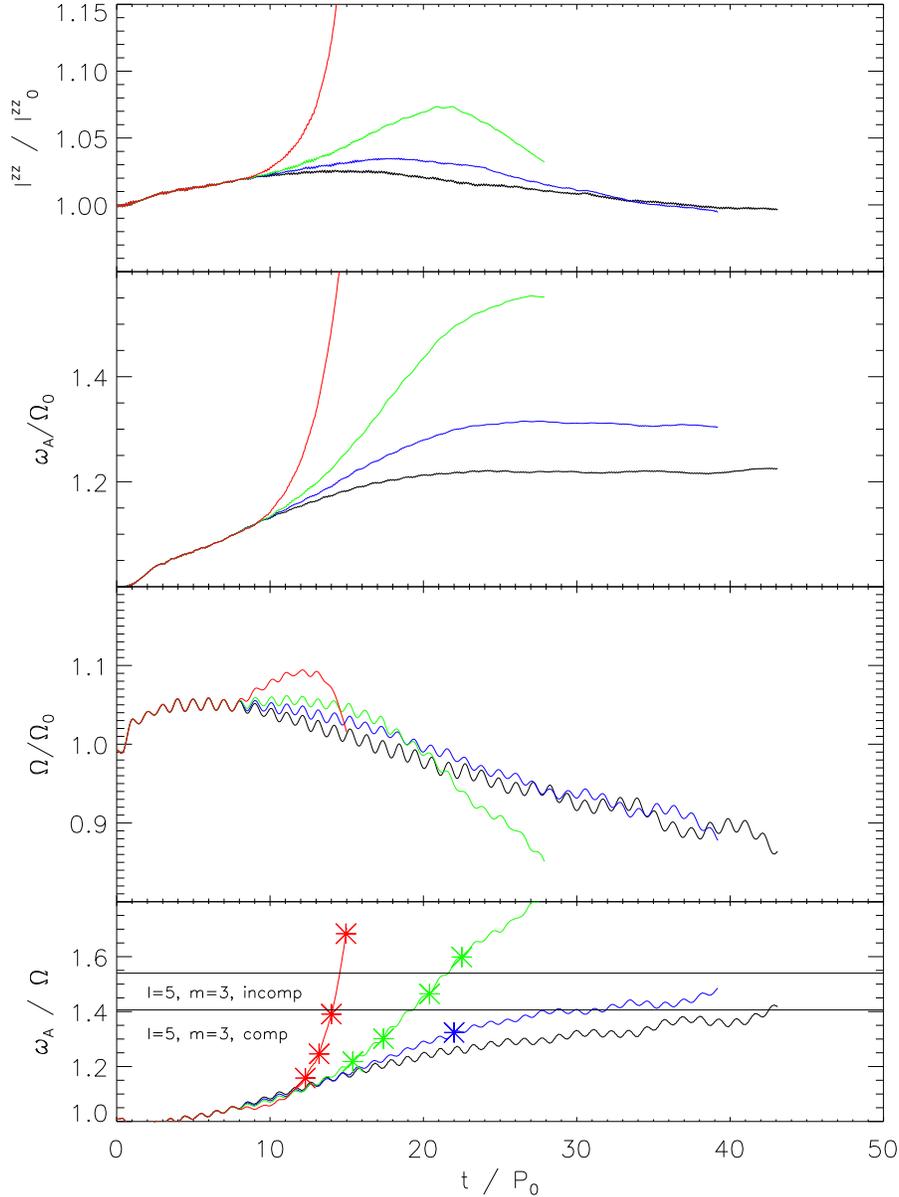}
   \end{center}
   \caption{From top to bottom: The moment of inertia $I_{zz}$ of the accretor around the spin axis normalized to its initial value,
   the mean spin angular velocity of the accretor $\omega_A$, the Keplerian estimate of the angular frequency of the binary
   $\Omega$, and the ratio $\omega_A/\Omega$ as functions of time.
   The angular velocities are referred to the inertial
   frame and are plotted normalized to the initial binary or corotating frame frequency. 
   The color coding for the curves is the same as in Fig.\ 4.
   The bottom frame also
   shows the two values corresponding to the resonant frequency ratios for the generalized $r$-mode with
   $l=5$ and $m=3$ for an incompressible spheroid and a polytrope of index $n=3/2$ as deduced from RPA. 
   The asterisks on the curves
   indicate the approximate times during which accretion belt disturbances resembling modes with
   (from left to right) $m=6$, 5, 4 and 3 are present (see the movies accompanying Fig. 3).
   Only $m=6$ is clearly discernible in run Q0.4B.}
   \label{omega_plots}
\end{figure}

 Clearly we would like to understand the mechanism whereby the binary averts its demise.
 Unfortunately we are not able at this point to offer an unambiguous interpretation.
 In an attempt to elucidate the observed behavior, we display in Fig. \ref{omega_plots}
 the evolution of the principal moment of inertia $I_{zz}$
 of the accretor, the mean spin angular velocity of the accretor $\omega_A = J_A/I_{zz}$,
 the angular velocity of the binary $\Omega=[G(M_A+M_B)/a^3]^{1/2}$, and the ratio of the latter two.
 The angular velocities are referred to the inertial frame and plotted normalized to
 $\Omega_{0}$. 
 In a recent preprint,
 \citet{RPA06} (hereafter RPA) discussed the possibility that a resonance condition between
 the orbital frequency and the eigenfrequencies of some of the generalized $r$-modes in the
 accretor star, saturates its spin
 and channels rotational kinetic energy into oscillation modes.
 This translates into a dissipationless torque capable of returning the spin angular
 momentum back to the orbit and thus increases the efficiency of tidal coupling.
 The fact that in our nonlinear simulations the change in the
 spin of the accretor is coupled to the dynamics of the binary, and that we observe equatorial distortions
 with $6\geq m \geq 3$ (see the movies accompanying Fig.\ 3), seem at first sight to suggest that these 
 modes play a role.
 However, when examined in detail, there are some aspects of the evolution that are inconsistent
 with the above interpretation.

 During the evolutions Q0.4A, B and C, the net torque on the accretor attains its largest
 positive (spin-up) value approximately at $t\approx 8$, 14 and 17, respectively (see Table 3)
 and the accretion rate peaks a few orbital periods later.
 Thus we know that the accretion torque is decreasing in magnitude, while the tidal torques
 trying to spin down the accretor are present and eventually the net torque changes sign
 at times $t\sim 20-25$.
 This indicates that the spin angular momentum is being
 returned to the orbit, the separation increases and the mass transfer rate continues to decline thereafter.
 In simulation Q0.4D, on the other hand, the driving is strong enough to keep the net torque positive
 throughout and the accretion rate increases monotonically until the donor is disrupted.
 A further interesting feature of the simulations is the appearance of distortions
 suggestive of modes with
 $3\leq m \leq 6$ at or after the spin-up torque peaks and before
  the time in which the spin saturates, while the orbital frequency
 continues to decline. 
 As has been indicated by the asterisks marking the curves in the bottom panel of Fig.\ 6,
 we observe in simulations Q0.4D and Q0.4C, and less clearly in
 Q0.4B, that equatorial distortions with $m=6$ appear first, and then evolve through
 decreasing $m=5,4$ and $m=3$. It is not clear whether even higher $m$ modes arise earlier
 on, nor if lower $m$ modes appear at the end. In every case these distortions arise during
 spin up and fade as the accretor spin levels off. Furthermore, the appearance of the
 azimuthal waves coincides roughly with the time in which the ratio
 $\omega_A/\Omega$ (even for run Q0.4D which fails to regain stability)
 passes through the range in which the
 resonance conditions for some, but not all of the modes are met.
 
 While the above features of our evolutions seem to suggest that the $r$-modes
 described by RPA are involved at some level,
 there are several reasons that limit the direct applicability of the
 results of RPA to our simulations: i) their calculated mode frequencies and
 estimated spin saturation frequencies are valid for an accretor in solid body rotation
 while the accretor in our simulations is differentially rotating and develops
 a prominent ``accretion belt"; ii) the extremely high accretion rate and stream impact
 in our simulations are significant deviations from the conditions assumed in RPA;
 iii) our simulations place no restrictions on the number, amplitude, or character of
 modes present, whereas RPA only consider generalized $r$-modes in the linear regime.
 In regard to this last point, although we have no direct measure of the effective polar
 quantum number $l$ dominant in the simulations, the thickness of the accretion belt
 does not appear to change much.
 Finally, while runs Q0.4A and B seem to saturate approximately at values
 $1.4\leq\omega_A/\Omega\leq 1.7$ (see bottom frame of Fig. \ref{omega_plots})
 consistent with the values predicted by RPA (1.54 for the incompressible case, and 1.406
 for a polytrope of index $n=3/2$, for a mode with $l=5$ and $m=3$), run Q0.4C levels off at
 a higher spin. Note that at late times, the accretion torque is significantly reduced
 as the accretion rate decreases, and that the leveling off of the spin is to be expected
 as the binary separates.
 In the scenario studied by RPA, the saturation frequency for the spin of the accretor 
 is a direct indication of the (single) generalized r-mode excited by the tidal gravitational 
 field.  In contrast, our simulations A through C all show different values for the 
 plateau in the accretor's spin frequency in Fig. 6 and these spin values scale with 
 increasing mass transfer rate.
 The diagnostic analysis that will be required to fully disentangle stream impact, tidal,
 and accretion effects unambiguously is beyond the scope of this paper.  A more detailed
 analysis of these interactions in the present simulations as well as in other mass
 transfer evolutions that we have performed is underway and will be reported elsewhere.

\clearpage
\clearpage

 To summarize, the picture that we have unveiled through these simulations is as
 follows.  Upon reaching contact, the binary is - at least initially - dynamically
 unstable to mass transfer.  This can be expressed with the inequality  that
 $q_{0} > q_{\rm crit}$.  The mass transfer rate grows rapidly but the binary then
 finds a means of escape.  The consequential angular momentum loss from the
 orbit becomes ineffective and angular momentum is returned to the orbit at nearly
 the same rate that it is advected through the accretion stream.  The binary must
 react as if $q_{\rm crit} = \frac{2}{3}$ and now $q(t) < q_{\rm crit}$.  The binary separation
 expands even faster, the mass transfer rate drops  and the dynamical instability
 is averted.

 \section{Discussion and Conclusion}

We have simulated, hydrodynamically, direct impact accretion in an initially semi-detached,
polytropic binary.  This model possesses an interesting mass ratio that is intermediate between
the two plausible stability bounds of $q_{\rm crit} = \frac{2}{3}$ for very efficient tidal coupling
and $q_{\rm crit} \sim 0.2$ that  applies when none of the advected angular momentum makes its
way back to the orbit.  From our baseline evolution we have demonstrated that for this
model, $q_{\rm crit} \geq 0.39$ as the mass transfer rate begins to decline once the binary has
crossed this mass ratio.

This lower bound for $q_{\rm crit}$ is refined further by evolving the model to its equilibrium
mass transfer rate corresponding to an imposed driving loss of angular momentum.  From
this numerical experiment, we find that $q_{\rm crit} \sim 0.7$, consistent with the expectation for
efficient tidal coupling in the binary.  This also indicates that the stability boundary itself
changes through the evolution.
By examining the simulations in light of the
individual terms in the orbit-averaged equation for
the binary separation (Equation \ref{oae}), we see that the contribution from the spin-up
of the accretor reaches an extreme value and then begins to decline in importance.
After the spin of the accretor begins to stabilize, the mass transfer rate begins to decline.
We therefore conclude that during the evolution, $q_{\rm crit}$ transitions from the tidally-inefficient
to tidally efficient limits on a timescale comparable to the mass-transfer time itself.
Although we observe that spin angular momentum is returned to the orbit by tidal coupling
to disturbances in the accretion belt, the detailed mechanism by which this is achieved
requires further investigation.

The behavior of the Q0.4 model supports the previous simulations we have performed for
a DWD binary with an initial mass ratio of $q = 0.5$ (DMTF).  In that work, we found that
the binary expanded and appeared likely to survive intact.  However, we were not able to
evolve the binary long enough to observe the mass transfer rate declining, though its growth
became progressively slower throughout that evolution.  We also note that for the DWD binary 
presented in DMTF, the initial binary components followed the expected mass-radius relation
for $n = \frac{3}{2}$ polytropes.
At this point, it is important to note that we have examined only two
binaries in detail to date and caution is warranted before generalizing our results.  This
note of caution is especially important for the results presented here for the Q0.4 model.
As discussed previously, the accretor is too large compared to the donor star to represent a
realistic DWD system and it is plausible that the relative size of the accretor may have
exaggerated the efficiency of torques within the binary.  To address this concern, we are
developing a more flexible SCF code to generate initial data that conform to prescribed
ratios for the mass and entropies of the two components.  Future simulations, for DWD models
with a range of initial mass ratios, will allow us to conclusively map out the effective stability
boundaries that hold for binaries undergoing direct impact accretion.

The effective stability bounds that have been discussed here considered only the hydrodynamic
flow of material in the self-consistent gravitational potential of the system.  In reality, we expect
that DWD binaries that begin dynamically unstable mass transfer - and even some systems
where the mass transfer remains stable (\citet{HW99}; GPF) - accrete at a rate in excess of the Eddington
rate.  Realistically, we can not neglect the influence of radiation on the dynamics of the accreted
material.  It is widely expected, though not conclusively demonstrated, that super-Eddington
mass transfer will fill a common envelope about the DWD causing the components to
inspiral and merge.

To put this statement in a concrete context,
in our simulations of the Q0.4 model, the lowest mass transfer rate we can accurately resolve
far exceeds the Eddington rate for a white dwarf accretor by 4-5 orders of magnitude.
To be consistent, we should really be examining accretion in DWD binaries as
a radiative hydrodynamics process.  While this goal lies well beyond the scope of the
current work, we feel that such simulations may uncover yet more unexpected behavior.
For example, the accretion flow from the donor will carry angular momentum that ensures
that, at least  in the initial phases of the mass transfer, the accreted material will be far from
a spherically symmetric state.  Radiation from the accretion shock may have a clear
channel to escape the system and a common envelope may not even form if the binary can
rapidly transition from dynamically unstable to stable mass transfer.

\acknowledgements
This work has been supported in part by NSF grants
AST 04-07070 and PHY 03-26311, and in part through NASA's ATP
program grants NAG5-8497, NAG5-13430 and NNX07AG84G.  The computations were
performed primarily at NCSA through grant MCA98N043, which allocated
resources on the Tungsten cluster, and on the SuperMike
cluster at LSU, which is provided by the Center for
Computation and Technology (CCT).

 \end{document}